\documentclass[journal]{IEEEtran}
\usepackage{cite}
\usepackage{graphicx}
\usepackage{amsmath}
\usepackage{times}
\usepackage{latexsym}
\usepackage{graphicx}
\usepackage{bm}
\usepackage{amssymb}
\usepackage[center]{caption2}
\usepackage{stfloats}
\usepackage{cases}
\usepackage{array}
\usepackage{setspace}
\usepackage{fancyhdr}
\usepackage{citesort}
\usepackage{color}
\usepackage{subfigure}

\renewcommand{\QED}{\QEDopen}
\newtheorem{theorem}{Theorem}

\newtheorem{corollary}{Corollary}
\newtheorem{proposition}{Proposition}

\def\proof{\noindent\hspace{2em}{\itshape Proof: }}
\def\endproof{\hspace*{\fill}~$\square$\par\endtrivlist\unskip}

\begin{document}

\title{Wireless Information and Power Transfer in Relay Systems with Multiple Antennas and Interference}
%
\author{Guangxu Zhu,~\IEEEmembership{Student Member,~IEEE,} Caijun Zhong,~\IEEEmembership{Senior Member,~IEEE,} Himal A. Suraweera,~\IEEEmembership{Member,~IEEE,} George K. Karagiannidis,~\IEEEmembership{Fellow,~IEEE,} Zhaoyang Zhang,~\IEEEmembership{Member,~IEEE} and Theodoros A. Tsiftsis,~\IEEEmembership{Senior Member,~IEEE}
\thanks{G. Zhu, C. Zhong and Z. Zhang are with the Institute of Information and Communication Engineering, Zhejiang University, China. (email: caijunzhong@zju.edu.cn).}
\thanks{H. A. Suraweera is with the Department of Electrical \& Electronic
Engineering, University of Peradeniya, Peradeniya 20400, Sri Lanka (email:
himal@ee.pdn.ac.lk).}
\thanks{G. K. Karagiannidis is with the Department of Electrical and Computer
Engineering, Khalifa University, PO Box 127788, Abu Dhabi, UAE and with the Department of Electrical and Computer
Engineering, Aristotle University of Thessaloniki, 54 124, Thessaloniki, Greece (e-mail: geokarag@ieee.org).}
\thanks{T. A. Tsiftsis is with the Department of Electrical Engineering, Technological Educational Institute of Central Greece, 35100 Lamia, Greece (email: tsiftsis@teilam.gr).}}
\maketitle

\begin{abstract}
In this paper, an energy harvesting dual-hop relaying system without/with the presence of co-channel interference
(CCI) is investigated. Specifically, the energy constrained mutli-antenna relay node is powered by either the
information signal of the source or via the signal receiving from both the source and interferer. In particular, we first
study the outage probability and ergodic capacity of an interference free system, and then extend the analysis to an
interfering environment. To exploit the benefit of multiple antennas, three different linear processing schemes are
investigated, namely, 1) Maximum ratio combining/maximal ratio transmission (MRC/MRT), 2) Zero-forcing/MRT
(ZF/MRT) and 3) Minimum mean-square error/MRT (MMSE/MRT). For all schemes, both the system¡¯s outage
probability and ergodic capacity are studied, and the achievable diversity order is also presented. In addition,
the optimal power splitting ratio minimizing the outage probability is characterized. Our results show that the
implementation of multiple antennas increases the energy harvesting capability, hence, significantly improves the
system¡¯s performance. Moreover, it is demonstrated that the CCI could be potentially exploited to substantially
boost the performance, while the choice of a linear processing scheme plays a critical role in determining how
much gain could be extracted from the CCI.


\end{abstract}



\section{Introduction}\label{section:1}
Energy harvesting technique, as an emerging solution for prolonging the lifetime of the energy constrained wireless devices, has gained significant interests in recent years. The conventional energy harvesting techniques rely on the external natural resources, such as solar power, wind energy or thermoelectric effects \cite{V.Raghunathan,B.Medepally,C.K.Ho}. However, due to the randomness and intermittent property of external natural resources, communication systems employing the conventional energy harvesting technique can not guarantee the delivery of reliable and uninterrupted communication services. Recently, the wireless energy transfer technique, first demonstrated by Nikola Tesla, has rekindled its interest in the context of energy harvesting communication systems where radio-frequency (RF) signals are used as an energy source \cite{Z.Ding1,I.Krikidis_2,L.R.Varshney,P.Grover,K.Huang}.
Since RF signals can be under control, it is much more reliable than external natural resources, hence, wireless energy harvesting using RF signals is a promising technique to power communication devices \cite{I.Krikidis}.

Since RF signals are capable of carrying both the information and energy, a new research area, namely {\it simultaneous wireless information
and power transfer} (SWIPT), has recently emerged. The seminal works \cite{L.R.Varshney,P.Grover} have characterized the fundamental tradeoff between the harvested energy and information capacity. Nevertheless, it was assumed in \cite{L.R.Varshney} that the receiver can decode the information and harvest energy from the same signal simultaneously, which is unfortunately impossible due to practical circuit limitations. To this end, the work in \cite{R.Zhang1} proposed two practical receiver architectures, namely, ``\emph{time-switching}'', where the receiver switches between decoding information and harvesting energy, and ``\emph{power-splitting}'', where the receiver splits the signal into two streams, one for information decoding and the other for energy harvesting. Since then, a number of works have appeared in the literature investigating different aspects of simultaneous information and energy transfer with practical receivers \cite{L.Liu,L.Liu2}. Specifically, in \cite{L.Liu}, an opportunistic RF energy harvesting scheme for single-input-single-output systems with co-channel interference (CCI) was investigated, where it was shown that the CCI can be identified as a potential energy source. Later on, an improved receiver, i.e., the dynamic power splitting receiver was studied in \cite{L.Liu2}. The extension of \cite{R.Zhang1} to the scenario with imperfect channel state information (CSI) at the transmitter was studied in \cite{M.Tao}. {For multiple-input single-output (MISO) channels, the optimal beamforming designs for SWIPT systems with/without secrecy constraint have been investigated in \cite{L.Liu3,J.Xu}, and the optimal transmission strategy maximizing the system throughput of MISO interference channel has been studied in \cite{C.Shen}. Moreover, the application of RF energy transfer technique in cognitive radio networks with multiple antennas at the secondary transmitter was considered in \cite{G.Zheng}.} Finally, cellular networks with RF energy transfer were considered in \cite{K.Huang,K.Huang2}. It is worth noting that all these prior works focus on the point-to-point communication systems.

The RF energy harvesting technique also finds important applications in cooperative relaying networks, where an energy-constrained relay with limited battery reserves relies on some external charging mechanism to assist the transmission of source information to the destination \cite{B.Medepally}. As such, a number of works have exploited the idea of achieving simultaneous information and energy transfer in cooperative relaying systems \cite{I.Krikidis0,A.Nasir,Z.Ding,Z.Ding1,I.Krikidis}. Specifically, \cite{A.Nasir} studied the throughput performance of an amplify-and-forward (AF) relaying system for both time-switching and power-splitting protocols and \cite{Z.Ding} considered the power allocation strategies for decode-and-forward (DF) relaying system with multiple source-destination pairs. More recently, the performance of energy harvesting cooperative networks with randomly distributed users was studied in \cite{Z.Ding1,I.Krikidis}. However, all these works are limited to the single antenna setup and all assume an interference free environment.

Motivated by this, we consider a dual-hop AF relaying system where the source and destination are equipped with a single antenna while the relay is equipped with multiple antennas.\footnote{{This particular system setup is applicable in several practical scenarios where two nodes (e.g., machine-to-machine type low cost devices) exchange information with the assistance of an advanced terminal such as a cellular base-station/clusterhead sensor.\cite{I.Krikidis_2,C.Zhong,G.Zhu}}} The energy constrained relay collects energy from ambient RF signals and uses the harvested energy to forward the information to the destination node. The power-splitting receiver architecture proposed in \cite{R.Zhang1} is adopted.
Specifically, we first study the performance of the multiple antenna relay system without CCI, which serves as a benchmark for the performance in the presence of CCI. Then, we present a detailed performance analysis for the system assuming a single dominant interferer at the relay. It is worth pointing out that, in the energy harvesting relaying system, while CCI provides additional energy, it corrupts the desired signal. In order to exploit CCI as a beneficial prospect, three different linear processing schemes, namely, 1) Maximum ratio combining/maximal ratio transmission (MRC/MRT), 2) Zero-forcing/MRT (ZF/MRT), 3) Minimum mean-square error/MRT (MMSE/MRT) are investigated.

The main contributions of this paper are summarized as follows:
\begin{itemize}
\item For the scenario without CCI, we derive an exact outage expression involving a single integral, and a tight closed-form outage probability lower bound. In addition, we present a simple high signal-to-noise ratio (SNR) approximation, which reveals that the system achieves a diversity order of $N$, where $N$ is the number of relay antennas. A new tight closed-form upper bound for the ergodic capacity is also derived. Finally, the optimal power splitting ratio minimizing the outage probability is characterized.

\item For the scenario with CCI, we present tight closed-form outage probability lower bounds and capacity upper bounds for all three schemes. In addition, we also characterize the high SNR outage behavior and show that both the MRC/MRT and MMSE/MRT schemes achieve a diversity order of $N$, while the ZF/MRT only achieves a diversity order of $N-1$. Moreover, the optimal power splitting ratio minimizing the outage probability is studied.

\item The presented analytical expressions provide an efficient means to evaluate key system performance metrics, such as the outage probability and ergodic capacity, without resorting to time-consuming Monte Carlo simulations. Therefore, a fast assessment of the impact of various key system parameters such as the energy harvesting efficiency $\eta$, the number of antennas $N$, the source transmitting power $\rho_1$ and the interference power $\rho_I$ on the optimal power splitting ratio is enabled.

\item Our results demonstrate that the CCI could be potentially exploited to significantly improve the system's performance. However, the actual performance gain due to CCI depends heavily on the choice of linear processing schemes. It is shown that the MMSE/MRT scheme is always capable of turning the CCI as a desired factor, and can achieve higher performance gain when the CCI is strong. On the other hand, CCI is not always beneficial when the MRC/MRT and ZF/MRT schemes are used. The performance degrades significantly in the strong CCI scenario if the MRC/MRT scheme is applied. In contrast, a weak interferer degrades the performance of the ZF/MRT scheme, which on the other hand achieves almost the same performance as the MMSE/MRT scheme in the presence of strong CCI.
\end{itemize}


The remainder of the paper is organized as follows: Section II introduces the system model. Section III investigates of the performance of the system without CCI. Section IV addresses the scenario with CCI. Numerical results and discussions are provided in Section V. Finally, Section VI concludes the paper and summarizes the key findings.

{\it Notation}: We use bold upper case letters to denote matrices, bold lower case letters to denote vectors and lower case
letters to denote scalars. ${\left\| {\bf{h}} \right\|_F}$ denotes the Frobenius norm; ${\tt E}\{x\}$ stands for the expectation of the random variable $x$; ${*}$ denotes the conjugate operator, while $T$ denotes the transpose operator and ${\dag}$ denotes the conjugate transpose operator; ${{\cal CN} (0,1)}$ denotes a scalar complex Gaussian distribution with zero mean and unit variance; $\Gamma(x)$ is the gamma function; $\Psi \left( {a,b;z} \right)$ is the confluent hypergeometric function \cite[Eq. (9.210.2)]{Tables}; $K_v(x)$ is the $v$-th order modified Bessel function of the second kind \cite[Eq. (8.407.1)]{Tables}; ${\rm Ei}(x)$ is the exponential integral function \cite[Eq. (8.211.1)]{Tables}; $\Gamma \left( {\alpha ,x} \right)$ is the upper incomplete gamma function \cite[Eq. (8.350.2)]{Tables}; ${}_2{F_1}(a,b;c;z)$ is the Gauss Hypergeometric function \cite[Eq. (9.100)]{Tables}; $\psi \left( x \right)$ is the Digamma function \cite[Eq. (8.360.1)]{Tables}; ${\mathop{\rm G}\nolimits}_{p,q}^{m,n} \left( \cdot\right)$ is the Meijer G-function \cite[Eq. (9.301)]{Tables} and ${\mathop{\rm G}\nolimits} _{1,[1:1],0,[1:1]}^{1,1,1,1,1}\left(  \cdot  \right)$ denotes the generalized Meijer G-function of two variables \cite{R.P.Agrawal} which can be computed by the algorithm presented in \cite[Table II]{I.S.Ansari}.

\section{System Model}
\begin{figure}[ht]
  \centering
  \subfigure[]{\label{fig:1a}\includegraphics[width=0.4\textwidth]{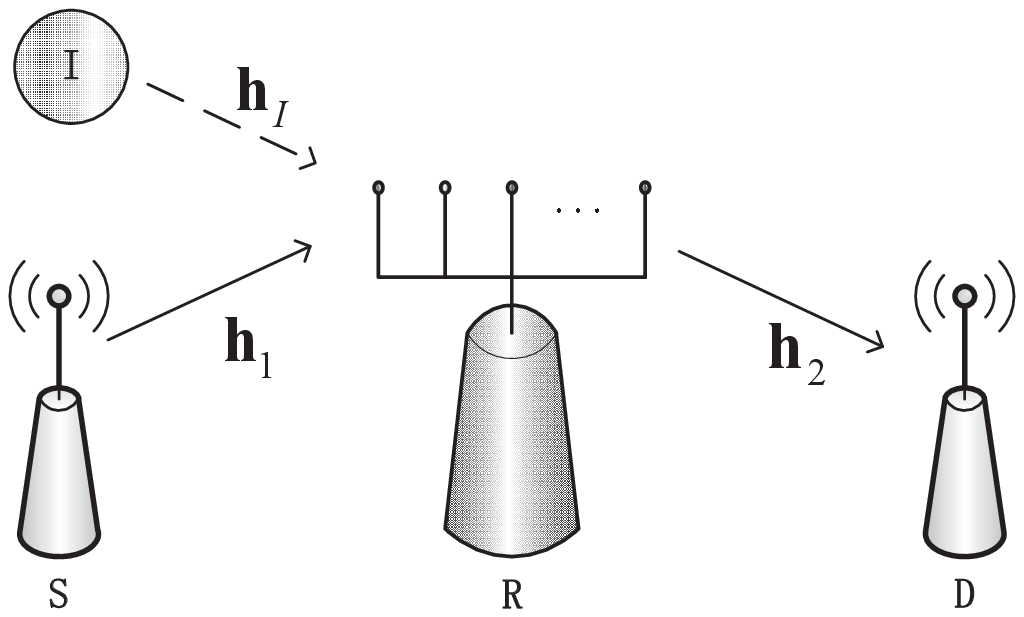}}
  \hspace{1in}
  \subfigure[]{\label{fig:1b}\includegraphics[width=0.4\textwidth]{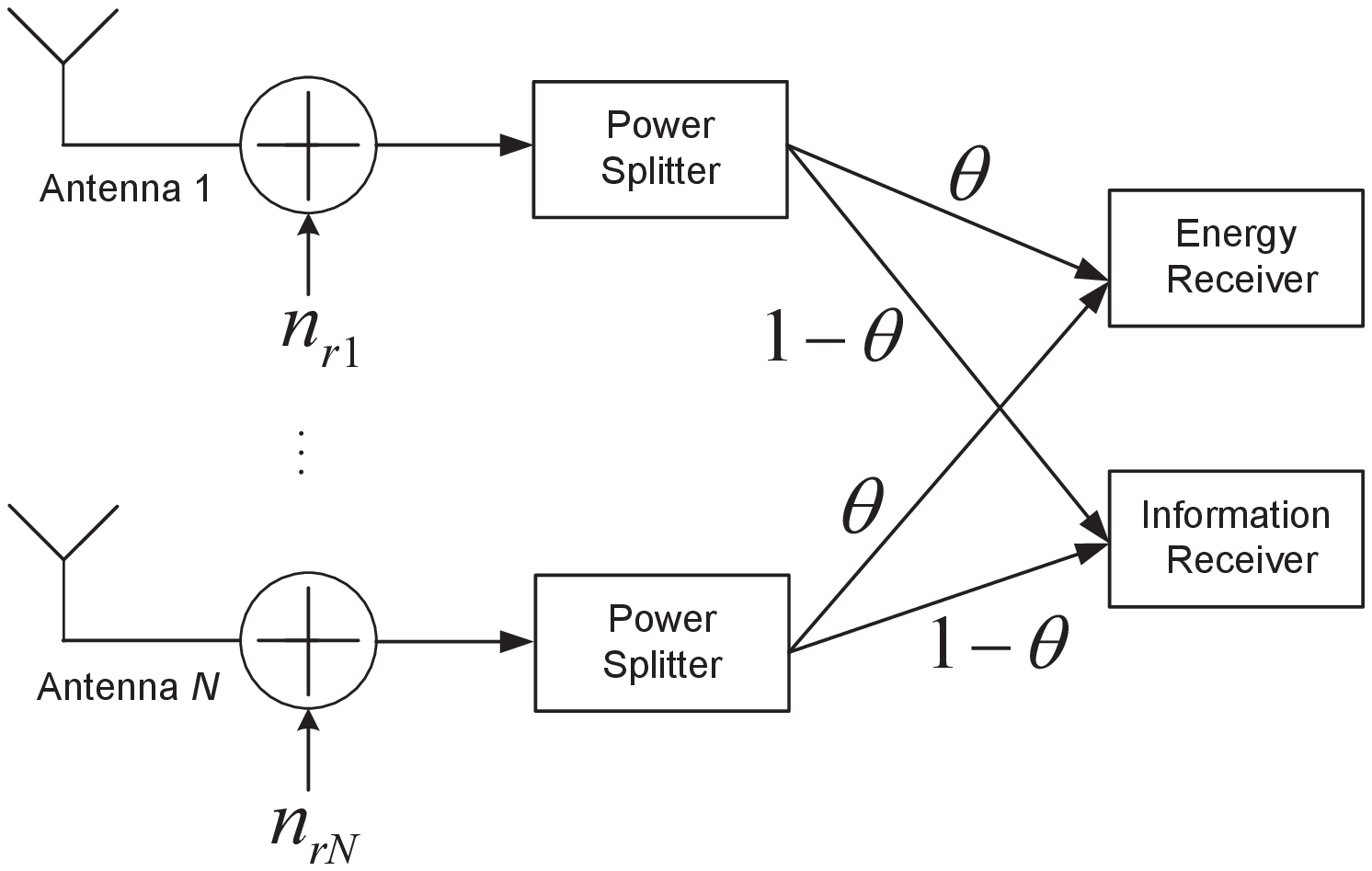}}
  \caption{(a) System model: S, R and D denote the source, relay and destination node, respectively. (b) Block diagram of the relay receiver with the power splitting protocol.}
  \label{fig:fig1}
\end{figure}

We consider a dual-hop multiple antenna AF energy harvesting relaying system as shown in Fig. \ref{fig:1a}, where both the source and the destination are equipped with a single antenna, while the relay is equipped with $N$ antennas \cite{I.Krikidis_2}. The source sends information to the destination through an energy constrained relay node. Throughout this paper, the following assumptions are adopted: 1) It is assumed that direct link between the source and the destination does not exist due to obstacles and/or severe fading. 2) The channel remains constant over the block time $T$ and varies independently and identically from one block to the other, and has a Rayleigh distributed magnitude. 3) As in \cite{X.Tang,C.Chae,R.Zhang}, no CSI is assumed at the source, full CSI is assumed at the relay, and local CSI is assumed at the destination.


We focus on the power splitting protocol proposed in \cite{R.Zhang1}. Specifically, the entire communication consists of two time slots with duration of $\frac{T}{2}$ each. At the end of the first phase,
each antenna at the relay node splits the received source signal into two
streams, one for energy harvesting and the other for information processing as depicted in Fig. \ref{fig:1b}. {As in \cite{Z.Ding,L.Liu2}, we consider the pessimistic case where power splitting only reduces the signal power, but not the noise power.} Hence, our results provide a lower bound on the performance for practical systems. We now consider two separate cases depending on whether the relay is subject to CCI or not.
\subsection{Noise-limited Case}
Let $\theta$ denote the power splitting ratio\footnote{{The optimality of uniform $\theta$ can be established by using similar methods as in [12].}}, then the signal component at the input of the information receiver is given by
\begin{align}\label{SM:1}
{{\bf{y}}_r} = \sqrt {{(1 - \theta ){P_s}}/{d_1^{\tau}}} {{\bf{h}}_1}x + {{\bf{n}}_r},
\end{align}
where $P_s$ denotes the source power, ${{{\bf h}}_{{1}}}$ is an ${N\times 1}$ vector with entries following identically and independently distributed (i.i.d.) ${{\cal CN} (0,1)}$, $d_1$ denotes the distance between the source and the relay, $\tau$ is the path loss exponent, $x$ is the source message with unit power, ${{{\bf n}}_{{r}}}$ is an ${N\times 1}$ vector and denotes the additive white Gaussian noise (AWGN) with ${\tt E}\{\mathbf{n}_r\mathbf{n}_r^{\dag}\} = N_0{\mathbf{I}}$.

At the end of the first phase, the overall energy harvested during half of the block time $\frac{T}{2}$, can be expressed as
\begin{align}
{E_h} = \frac{\eta \theta {P_s}}{d_1^{\tau}}\left\| {{{\bf{h}}_1}} \right\|_F^2\frac{T}{2},\end{align}
where $\eta$ denotes the RF-to-DC conversion efficiency.

%
\subsection{Interference plus Noise Case}
We assume that the relay is subjected to a single dominant interferer and AWGN while the destination is still corrupted by the AWGN only.\footnote{{The scenario where the relay and the destination experience different interference patterns will occur in frequency-division relaying systems \cite{R.Pabst}.}} It is worth pointing out the single dominant interferer assumption has been widely adopt in the literature, see \cite{R.Mallik,M.Hassanien} and references therein. Moreover, such a system model enables us to gain key insights on the joint effect of CCI and multiple antennas in an energy harvesting relaying system.

In such case, the signal at the input of the information receiver at the relay is given by
\begin{align}\label{ESM:2}
{{\bf{y}}_r} = \sqrt {{(1 - \theta ){P_s}}/{d_1^{\tau}}} {{\bf{h}}_1}x + \sqrt {{(1 - \theta ){P_I}}/{{d_I^{\tau}}}} {{\bf{h}}_I}{s_I} + {{\bf{n}}_r},
\end{align}
where $P_I$ is the interference power, $d_I$ denotes the distance between the interferer and the relay, ${s_I}$ is the interference symbol with unit power, and ${{{\bf h}}_{{I}}}$ is an ${N\times 1}$ vector with entries following i.i.d. ${{\cal CN} (0,1)}$.

Also, according to \cite{L.Liu}, at the end of the first phase, the overall energy harvested during half of the block time $\frac{T}{2}$ is given by
\begin{align}
{E_h} = \eta \theta \left( {\frac{P_s}{d_1^{\tau}}\left\| {{{\bf{h}}_1}} \right\|_F^2 + \frac{P_I}{d_I^{\tau}}\left\| {{{\bf{h}}_I}} \right\|_F^2} \right)\frac{T}{2}.
\end{align}

For both cases, during the second phase, the relay transmits a transformed version of the received signal to the destination using the harvested power. Hence, the signal at the destination can be expressed as
\begin{align}\label{SM:3}
{y_d} = \sqrt{{1}/{d_2^{\tau}}}{{\bf{h}}_2}{\bf{W}}{{\bf{y}}_r} + {n_d},
\end{align}
where ${{\bf{h}}_{{2}}}$ is a ${1\times N}$ vector and denotes the relay-destination channel and its entries follow i.i.d. ${{\cal CN} (0,1)}$, $d_2$ denotes the distance between the relay and the destination, ${n_d}$ is the AWGN at the destination with ${\tt E}\{n_d^{*}n_d\} = N_0$, $\mathbf{W}$ is the transformation matrix applied at the information receiver at the relay with ${\tt E}\{\left\|\mathbf{W}\mathbf{y}_r\right\|_F^2\} = P_r$. Obviously, the performance of the system depends on the choice of $\mathbf{W}$, which will be elaborated in the ensuing sections.

\section{The noise-limited Scenario}
In this section, we consider the scenario where the relay is corrupted by AWGN only. In such case, it can be shown that the optimal transformation matrix $\bf{W}$ has the following structure:
\begin{align}\label{SM:4}
{\bf{W}} = \omega \frac{{{\bf{h}}_2^{\bf{\dag }}}{{\bf{h}}_1^{\bf{\dag }}}}{{\left\| {{{\bf{h}}_2}} \right\|}_F{\left\| {{{\bf{h}}_1}} \right\|}_F},
\end{align}
where $\omega$ is the power constraint factor, i.e., the information receiver first applies the MRC principle to combine all the signals from $N$ antennas, and then forward the signal to the destination by using the MRT principle. To guarantee the transmit power constraint at the relay, $\omega$ can be computed as
\begin{align}\label{SM:5}
{\omega ^2} = \frac{{P_r}}{{\frac{\left( {1 - \theta } \right){P_s}}{d_1^{\tau}}\left\| {{{\bf{h}}_1}} \right\|_F^2 + {N_0}}},
\end{align}
where $P_r$ is the available relay power. Since the relay communicates with the destination for half of the block time $\frac{T}{2}$, we have ${P_r} = \frac{{{E_h}}}{{{T \mathord{\left/
 {\vphantom {T 2}} \right.
 \kern-\nulldelimiterspace} 2}}} = \frac{\eta \theta {P_s}}{d_1^{\tau}}\left\| {{{\bf{h}}_1}} \right\|_F^2.$
Hence, the end-to-end SNR of the system can be expressed as
\begin{align}\label{SM:7}
\gamma  &= \frac{{{\omega ^2}\left\| {{{\bf{h}}_2}} \right\|_F^2\left\| {{{\bf{h}}_1}} \right\|_F^2\frac{\left( {1 - \theta } \right){P_s}}{d_1^{\tau}d_2^{\tau}}}}{{{\omega ^2}\left\| {{{\bf{h}}_2}} \right\|_F^2\frac{N_0}{d_2^{\tau}} + {N_0}}} \notag\\
&= \frac{{\frac{\eta \theta \left( {1 - \theta } \right)\rho _1^2}{d_1^{2\tau}d_2^{\tau}}\left\| {{{\bf{h}}_2}} \right\|_F^2\left\| {{{\bf{h}}_1}} \right\|_F^4}}{{\frac{\eta \theta {\rho _1}}{{d_1^{\tau}d_2^{\tau}}}\left\| {{{\bf{h}}_2}} \right\|_F^2\left\| {{{\bf{h}}_1}} \right\|_F^2 + \frac{\left( {1 - \theta } \right){\rho _1}}{d_1^{\tau}}\left\| {{{\bf{h}}_1}} \right\|_F^2 + 1}},
\end{align}
where $\rho_1$ is defined as ${\rho _1} = {{P_s}}/{{{N_0}}}$.

In the following, we give a detailed performance analysis in terms of the outage probability and ergodic capacity. In addition, the optimal $\theta$ minimizing the outage probability is investigated.
\subsection{Outage Probability}\label{sec:1}
The outage probability is an important performance metric, which is defined as the instantaneous SNR falls below a pre-defined threshold $\gamma_{\sf th}$. Mathematically, outage probability can be expressed as
\begin{align}\label{opa:1}
{P_{\sf out}} = {\mathop{\rm Prob}\nolimits} \left( {\gamma  < {\gamma _{\sf th}}} \right).
\end{align}
\begin{theorem}\label{theorem:1}
The outage probability of the multiple antenna energy harvesting relaying system can be expressed as
\begin{align}\label{opa:2}
{P_{{\sf{out}}}} = 1 - \int_{d/c}^\infty  {\frac{{\Gamma \left( {N,\frac{{ax + b}}{{c{x^2} - dx}}} \right)}}{{\Gamma \left( N \right)}}\frac{{{x^{N - 1}}}}{{\Gamma \left( N \right)}}{e^{ - x}}dx},
\end{align}

where $a = \frac{\left( {1 - \theta } \right){\rho _1}{\gamma _{{\sf{th}}}}}{d_1^{\tau}}$, $b = {\gamma _{{\sf{th}}}}$, $c = \frac{\eta \theta \left( {1 - \theta } \right)\rho _1^2}{d_1^{2\tau}d_2^{\tau}}$, $d = \frac{\eta \theta {\rho _1}{\gamma _{{\sf{th}}}}}{d_1^{\tau}d_2^{\tau}}$.

\proof
Substituting (\ref{SM:7}) into (\ref{opa:1}), the outage probability of the system can be expressed as
\begin{align}\label{Aopa:1}
&{P_{{\rm{out}}}} = \\
&{\rm{Prob}}\left( {\left\| {{{\bf{h}}_2}} \right\|_F^2\left( {c\left\| {{{\bf{h}}_1}} \right\|_F^4 - d\left\| {{{\bf{h}}_1}} \right\|_F^2} \right) < \left( {a\left\| {{{\bf{h}}_1}} \right\|_F^2 + b} \right)} \right),\notag
\end{align}
which can be evaluated as
\begin{multline}\label{Aopa:4}
{P_{{\rm{out}}}} =\int_0^{d/c} {{f_{\left\| {{{\bf{h}}_1}} \right\|_F^2}}\left( x \right)dx}  \\+ \int_{d/c}^\infty  {{f_{\left\| {{{\bf{h}}_1}} \right\|_F^2}}\left( x \right){F_{\left\| {{{\bf{h}}_2}} \right\|_F^2}}\left( {\frac{{ax + b}}{{c{x^2} - dx}}} \right)dx}.
\end{multline}
%
Since the squared Frobenius norm of a complex Gaussian vector is Chi-square distributed, $\left\| {{{\bf{h}}_1}} \right\|_F^2$ and $\left\| {{{\bf{h}}_2}} \right\|_F^2$ are i.i.d. gamma random variables. After some simple algebraic manipulations (\ref{opa:2}) is obtained.
%
\endproof
\end{theorem}

Theorem \ref{theorem:1} presents the exact outage probability of the of the system with arbitrary number of antennas. For the special case with a single antenna at the relay, Theorem \ref{theorem:1} reduces to the result derived in \cite[Proposition 3]{A.Nasir}.
To the best of the authors' knowledge, the integral in (\ref{opa:2}) does not admit a closed-form expression. However, this
single integral expression can be efficiently evaluated numerically using software such as Matlab or MATHEMATICA. Alternatively, we can use the following closed-form lower bound for the outage probability, {which will be shown to be tight across the entire SNR range in the Section \ref{sec:5}.}

\begin{corollary}\label{coro:1}
The outage probability of the multiple antenna energy harvesting relaying system can be lower bounded as
\begin{multline}\label{opa:3}
P_{{\sf{out}}}^{{\sf{low}}} = 1 - \frac{{2{e^{ - d/c}}}}{{\Gamma \left( N \right)}}\sum\limits_{i = 0}^{N - 1} {\frac{1}{{i!}}\sum\limits_{j = 0}^{N - 1} {{N-1}\choose j} } {\left( {\frac{d}{c}} \right)^{N - j - 1}}\times\\{\left( {\frac{a}{c}} \right)^{\frac{{i + j + 1}}{2}}}{K_{i - j - 1}}\left( {2\sqrt {\frac{a}{c}} } \right).
\end{multline}

\proof See Appendix \ref{appendix:corollary:1}. \endproof
\end{corollary}

While Theorem \ref{theorem:1} and Corollary \ref{coro:1} are useful to study the system's outage probability, the expressions are in general too complex to gain insight. Motivated by this, we now look into the high SNR regime, and derive a simple approximation for the outage probability, which enables the characterization of the achievable diversity order.

\begin{theorem}\label{theorem:2}
In the high SNR regime, i.e., ${\rho _1} \to \infty $, the outage probability of the multiple antenna energy harvesting relaying system can be approximated as
\begin{multline}\label{opa:4}
P_{{\sf{out}}}^\infty  \approx \frac{\left( {\frac{{{d_1^{\tau}\gamma _{{\sf{th}}}}}}{{{\rho _1}}}} \right)^N}{{\Gamma \left( {N + 1} \right)}}\left( {\frac{1}{{{{\left( {1 - \theta } \right)}^N}}}}\right. +\\
\left.{ \frac{{\ln \left( {\left( {1 - \theta } \right){\rho _1}} \right) - \ln {(d_1^{\tau}\gamma _{{\sf{th}}})} - {\bf{C}}}}{{\Gamma \left( N \right)}}{{\left( {\frac{d_2^{\tau}}{{\eta \theta }}} \right)}^N}} \right),
\end{multline}
where $\bf{C}$ is the Euler-Mascheroni constant \cite[Eq. (9.73)]{Tables}.

\proof See Appendix \ref{appendix:theorem:2}. \endproof
\end{theorem}

We observe that the system achieves a diversity order of $N$, which is the same as the conventional case with constant power relay node \cite{C.Zhong}. However, we notice that $P_{\sf out}$ decays as $\rho_1^{-N}\ln {\rho_1}$ rather than $\rho_1^{-N}$ as in the conventional case \cite{C.Zhong}. This important observation implies that, in the energy harvesting case, the slope of $P_{\sf out}$ converges much slower compared with that in the constant power case. {Please note that similar observations have been made in prior work [21]. The possible reason is that, in SWIPT systems, the available transmit power at the relay is a random variable, which results in higher outage probability compared to the conventional constant relay power case.}

\subsection{Ergodic Capacity}\label{sec:2}
%
Noticing that the end-to-end SNR given in (\ref{SM:7}) can be alternatively expressed as
\begin{align}\label{eca:2}
\gamma  = \frac{{{\gamma _1}{\gamma _2}}}{{{\gamma _1} + {\gamma _2} + 1}},
\end{align}
where ${\gamma _1} = \frac{\left( {1 - \theta } \right){\rho _1}}{d_1^{\tau}}\left\| {{{\bf{h}}_1}} \right\|_F^2$ and ${\gamma _2} = \frac{\eta \theta {\rho _1}}{d_1^{\tau}d_2^{\tau}}\left\| {{{\bf{h}}_2}} \right\|_F^2\left\| {{{\bf{h}}_1}} \right\|_F^2$. The ergodic capacity is given by
\begin{align}\label{eca:3}
C = \frac{1}{2}{\mathop{\rm E}\nolimits} \left[ {{{\log }_2}\left( {1 + \frac{{{\gamma _1}{\gamma _2}}}{{{\gamma _1} + {\gamma _2} + 1}}} \right)} \right].
\end{align}
Unfortunately, an exact evaluation of the ergodic capacity is generally intractable, since the cumulative distribution function (c.d.f.) of (\ref{SM:7}) can not be given in closed-form. Motivated
by this, we hereafter seek to deduce a tight bound for the ergodic capacity.

Starting from (\ref{eca:3}), the ergodic capacity can be alternatively expressed as
\begin{align}\label{eca:4}
C = \frac{1}{2}{\mathop{\rm E}\nolimits} \left[ {{{\log }_2}\left( {\frac{{\left( {1 + {\gamma _1}} \right)\left( {1 + {\gamma _2}} \right)}}{{1 + {\gamma _1} + {\gamma _2}}}} \right)} \right]= C_{\gamma _1}+C_{\gamma _2}-C_{\gamma _T},
\end{align}
where $C_{\gamma _i} = \frac{1}{2}{\mathop{\rm E}\nolimits} \left[ {{{\log }_2}\left( {1 + {\gamma _i}} \right)} \right]$, for $i \in \{ 1,2\} $, and $C_{\gamma _T} = \frac{1}{2}{\mathop{\rm E}\nolimits} \left[ {{{\log }_2}\left( {1 + {\gamma _1}+ {\gamma _2}} \right)} \right]$. Using the fact that $f\left( {x,y} \right) = {\log _2}\left( {1 + {e^x} + {e^y}} \right)$ is a convex function with respect to $x$ and $y$, we have
\begin{align}\label{eca:5}
{C_{{\gamma _T}}} \ge \frac{1}{2}{\log _2}\left( {1 + {e^{{\mathop{\rm E}\nolimits} \left( {\ln {\gamma _1}} \right)}} + {e^{{\mathop{\rm E}\nolimits} \left( {\ln {\gamma _2}} \right)}}} \right).
\end{align}
We now establish the ergodic capacity upper bound of the system using the following theorem:
\begin{theorem}\label{theorem:3}
The ergodic capacity of the multiple antenna energy harvesting relaying system is upper bounded by
\begin{multline}\label{eca:6}
{C_{\sf up}} = \frac{e^{\frac{d_1^{\tau}}{{\left( {1 - \theta } \right){\rho _1}}}}}{{2\ln 2}}\sum\limits_{k = 0}^{N - 1} {{{\left( {\frac{d_1^{\tau}}{{\left( {1 - \theta } \right){\rho _1}}}} \right)}^k}\Gamma \left( { - k,\frac{d_1^{\tau}}{{\left( {1 - \theta } \right){\rho _1}}}} \right)} +\\
\frac{1}{{2\ln 2}}\frac{1}{{\Gamma \left( {\rm{N}} \right)}}\sum\limits_{m = 0}^{N - 1} \frac{1}{{m!}}{{\left( {\frac{d_1^{\tau}d_2^{\tau}}{{\eta \theta {\rho _1}}}} \right)}^m} G_{1,3}^{3,1}\left( \frac{d_1^{\tau}d_2^{\tau}}{{\eta \theta {\rho _1}}}\middle|^
{ - m}_
{ - m,N - m,0}
\right) -\\ \frac{1}{2}{\log _2}\left( {1 + \frac{(1 - \theta){\rho _1}}{d_1^{\tau}}{e^{\psi \left( N \right)}} + \frac{\eta \theta {\rho _1}}{d_1^{\tau}d_2^{\tau}}{e^{2\psi \left( N \right)}}} \right).
\end{multline}
\proof See Appendix \ref{appendix:theorem:3}. \endproof
\end{theorem}

Theorem \ref{theorem:3} presents a new upper bound for the ergodic capacity of the system, which is quite tight across the entire SNR range as shown in the section \ref{sec:5}, hence, providing an efficient means to evaluate the ergodic capacity without resorting to Monte Carlo simulations. In addition, as we show in the next subsection, it enables the study of the optimal power splitting ratio.

\subsection{Optimization of the Parameter $\theta$ in High SNR Value}\label{sec:3.3}
The right selection of the power splitting ratio $\theta$ is crucial for the system's performance. A high value of $\theta$ could provide more transmission power at the relay, which benefits the relay-destination transmission. Nevertheless, a large $\theta$ also deteriorates the quality of the source-relay transmission. Hence, there exists a delicate balance, which we now investigate. For tractability, we only focus on the outage performance {in the high SNR region}, and the impact of $\theta$ on the ergodic capacity will be numerically illustrated in the Section \ref{sec:5}.


Starting from the high SNR approximation of ${P_{\sf out}}$ in (\ref{opa:4}), the optimal $\theta$, which is the solution of the optimization problem $\mathop {\min }\limits_{0 < \theta  < 1} \;{P_{\sf out}}$, can be obtained by solving the equivalent problem in (\ref{ota:2}) shown on the top of the next page.
\begin{figure*}
\begin{align}\label{ota:2}
\mathop {\min }\limits_{0 < \theta  < 1} \;\;f\left( \theta  \right) = \frac{1}{{{{\left( {1 - \theta } \right)}^N}}} + \frac{{\ln \left( {\left( {1 - \theta } \right){\rho _1}} \right) - \ln {d_1^{\tau}\gamma _{{\rm{th}}}} - {\bf{C}}}}{{\Gamma \left( N \right)}}{\left( {\frac{d_2^{\tau}}{{\eta \theta }}} \right)^N}.
\end{align}
\hrule
\end{figure*}

\begin{proposition}\label{prop:1}
The optimal $\theta$ is the root of the following polynomial
\begin{multline}\label{ota:3}
{a_1}{\theta ^{N + 1}} - {b_1}{\left( {1 - \theta } \right)^{N + 1}} - {c_1}\theta {\left( {1 - \theta } \right)^N} \\- {d_1}{\left( {1 - \theta } \right)^{N + 1}}\ln \left( {1 - \theta } \right) = 0,
\end{multline}
where ${a_1} = N$, ${b_1} = \frac{d_2^{N\tau}N\left( {\ln {\rho _1} - \ln {d_1^{\tau}\gamma _{{\rm{th}}}} - {\bf{C}}} \right)}{{{\eta ^N}\Gamma \left( N \right)}}$, ${c_1} = \frac{d_2^{N\tau}}{{{\eta ^N}\Gamma \left( N \right)}}$, ${d_1} = \frac{Nd_2^{N\tau}}{{{\eta ^N}\Gamma \left( N \right)}}$ and $0<\theta<1$ .

\proof
{It is easily to prove that, when ${\rho _1} \to \infty $, there is only one root (denoted by $\theta^*$) on the interval of $(0,1)$ for the equation $f'(\theta)=0$, and we can also note that $f'(0)=-\infty$ and $f'(1)=+\infty$. Due to the continuity of $f'(\theta)$, we have $f'(\theta) < 0$, $\theta \in (0,\theta^*)$ and $f'(\theta) > 0$, $\theta \in (\theta^*,1)$, which means that $f(\theta)$ first decreases as $\theta$ from $0$ to $\theta^*$ and then increases as $\theta$ from  $\theta^*$ to $1$. Therefore, the global minimum of $f(\theta)$ can be obtained by solving $f'\left( \theta  \right) = 0$.}
%
\endproof
\end{proposition}

Due to the presence of logarithmic function $\ln{(1-\theta)}$, a closed-form expression for the root of (\ref{ota:3}) can not be obtained. However, it can be efficiently solved numerically.

\section{The Interference plus Noise Scenario}
We now assume that the relay is subject to the influence of a single dominant interferer. {In the presence of CCI, the optimal relay processing matrix ${\bf W}$ maximizing the end-to-end signal-to-interference-and-noise ratio (SINR) of the system is the solution of the following optimization problem:
\begin{align}
&\mathop {\max }\limits_{\bf{W}} \;\gamma  = \frac{{\frac{{(1 - \theta ){P_s}}}{{d_1^\tau d_2^\tau }}{{\left| {{{\bf{h}}_2}{\bf{W}}{{\bf{h}}_1}} \right|}^2}}}{{\frac{{(1 - \theta ){P_I}}}{{d_1^\tau d_2^\tau }}{{\left| {{{\bf{h}}_2}{\bf{W}}{{\bf{h}}_I}} \right|}^2} + \frac{{\left\| {{{\bf{h}}_2}{\bf{W}}} \right\|_F^2}}{{d_2^\tau }}{N_0} + {N_0}}}\notag\\
&s.t.\;\;{\tt E}\{ \left\| {{\bf{W}}{{\bf{y}}_r}} \right\|_F^2\}  = {P_r} = \eta \theta \left( {\frac{{{P_s}}}{{d_1^\tau }}\left\| {{{\bf{h}}_1}} \right\|_F^2 + \frac{{{P_I}}}{{d_I^\tau }}\left\| {{{\bf{h}}_I}} \right\|_F^2} \right).
\end{align}
Due to the non-convex nature of the problem, a closed-form solution for ${\bf W}$ is hard to find. Hence, in the following, we consider three heuristic two-stage relay processing strategies proposed in \cite{G.Zhu}, i.e., the matrix ${\bf W}$ admits the rank-1 structure  ${\bf{W}} = \omega \frac{{{\bf{h}}_2^{\bf{\dag }}}}{{\left\| {{{\bf{h}}_2}} \right\|}_F}{{\bf{w}}_1}$, where ${{\bf{w}}_1}$ is a ${1\times N}$ linear combining vector, which depends on the linear combining scheme employed at the relay and will be specified in the following subsection.}

\subsection{MRC/MRT Scheme}
For the MRC/MRT scheme, ${{\bf{w}}_1}$ is set to match the first hop channel given in (\ref{SM:4}).
To meet the transmit power constraint at the relay, the power constraint factor $\omega^2$ should be given by
\begin{align}\label{ESM:3}
{\omega ^2} = \frac{{{P_r}}}{{\frac{\left( {1 - \theta } \right){P_s}}{d_1^{\tau}}\left\| {{{\bf{h}}_1}} \right\|_F^2 + \frac{\left( {1 - \theta } \right){P_I}}{d_I^{\tau}}\frac{{{{\left| {{\bf{h}}_1^\dag {{\bf{h}}_I}} \right|}^2}}}{{\left\| {{{\bf{h}}_1}} \right\|_F^2}} + {N_0}}},
\end{align}
where ${P_r} = \frac{{{E_h}}}{{{T \mathord{\left/
 {\vphantom {T 2}} \right.
 \kern-\nulldelimiterspace} 2}}} = \eta \theta \left( {\frac{P_s}{d_1^{\tau}}\left\| {{{\bf{h}}_1}} \right\|_F^2 + \frac{P_I}{d_I^{\tau}}\left\| {{{\bf{h}}_I}} \right\|_F^2} \right)$.
Therefore, the end-to-end SINR of the MRC/MRT scheme can be expressed as
\begin{align}\label{ESM:4}
{\gamma _I^{\sf MRC}} = \frac{{{\gamma _{I1}^{\sf MRC}}{\gamma _{I2}^{\sf MRC}}}}{{{\gamma _{I1}^{\sf MRC}} + {\gamma _{I2}^{\sf MRC}} + 1}},
\end{align}
where ${\gamma _{I1}^{\sf MRC}} = \frac{{\frac{\left( {1 - \theta } \right){\rho _1}}{d_1^{\tau}}\left\| {{{\bf{h}}_1}} \right\|_F^2}}{{\frac{\left( {1 - \theta } \right){\rho _I}}{d_I^{\tau}}\frac{{{{\left| {{\bf{h}}_1^\dag {{\bf{h}}_I}} \right|}^2}}}{{\left\| {{{\bf{h}}_1}} \right\|_F^2}} + 1}}$, ${\gamma _{I2}^{\sf MRC}} = \frac{\eta \theta}{d_2^{\tau}} \left( {\frac{\rho _1}{d_1^{\tau}}\left\| {{{\bf{h}}_1}} \right\|_F^2 + \frac{\rho _I}{d_I^{\tau}}\left\| {{{\bf{h}}_I}} \right\|_F^2} \right)\left\| {{{\bf{h}}_2}} \right\|_F^2$ and $\rho_I$ is defined as ${\rho _I} = \frac{{P_I}}{{{N_0}}}$.

\subsubsection{Outage Probability}
Since the exact analysis appears to be difficult, in the following we focus on deriving an outage lower bound and a simple high SNR outage approximation. {According to \cite{G.Zhu,C.Zhong}}, the end-to-end SINR in (\ref{ESM:4}) can be tightly upper bounded by
\begin{align}\label{EOP:1}
{\gamma _I^{\sf MRC}} \le {\gamma _I^{\sf up}} = \min \left( {{\gamma _{I1}^{\sf MRC}}\;,\;{\gamma _{I2}^{\sf MRC}}} \right),
\end{align}
the outage probability of the MRC/MRT scheme is lower bounded by
\begin{align}\label{EOP:2}
{P_{{I\sf{out}}}^{\sf LMRC}} =  {\rm{Prob}}\left( {\gamma _I^{\sf up} < {\gamma _{{\sf{th}}}}} \right).
\end{align}
We have the following key result:
\begin{theorem}\label{theorem:4}
If ${\rho _1} \ne {\rho _I}$,\footnote{For mathematical tractability, we only provide the result for the general case where the signal from the source and the CCI have different power, i.e., ${\rho _1} \ne {\rho _I}$. But the result for the special case ${\rho _1} = {\rho _I}$ is much more easier and can be obtained in a similar way.} the outage probability of the MRC/MRT scheme can be lower bounded as
\begin{align}\label{EOP:3}
{P_{{I\sf{out}}}^{\sf LMRC}} = 1 - {F_1^{\sf MRC}}{F_2^{\sf MRC}},
\end{align}
with
\begin{align}
{F_1^{\sf MRC}} = \frac{{{d_I^{\tau}e^{ - \frac{{{d_1^{\tau}\gamma _{{\sf{th}}}}}}{{\left( {1 - \theta } \right){\rho _1}}}}}}}{{\left( {1 - \theta } \right){\rho _I}}}{\sum\limits_{m = 0}^{N - 1} {\left( {\frac{{{d_1^{\tau}\gamma _{{\sf{th}}}}}}{{\left( {1 - \theta } \right){\rho _1}}}} \right)} ^m}\times \notag\\
\sum\limits_{n = 0}^m {\frac{1}{{(m - n)!}}{{\left( {\frac{{\left( {1 - \theta } \right){\rho _1}{\rho _I}}}{{{d_I^{\tau}\rho _1} + {d_1^{\tau}\rho _I}{\gamma _{{\sf{th}}}}}}} \right)}^{n + 1}}},\notag
\end{align}
and ${F_2^{\sf MRC}}$ can be expressed as in (\ref{EOP:5}) shown on the top of the next page.
\begin{figure*}
\begin{multline}\label{EOP:5}
{F_2^{\sf MRC}} = \frac{2d_1^{N\tau}d_I^{N\tau}}{{\rho _1^N\rho _I^N}}\sum\limits_{s = 1}^N {\frac{{\prod\nolimits_{j = 1}^{s - 1} {\left( {1 - N - j} \right)} }}{{\left( {N - s} \right)!\left( {s - 1} \right)!}}{{\left( {\frac{d_I^{\tau}}{{{\rho _I}}} - \frac{d_1^{\tau}}{{{\rho _1}}}} \right)}^{1 - N - s}}}
\sum\limits_{m = 0}^{N - 1} \frac{1}{{m!}}{{\left( {\frac{{{d_2^{\tau}\gamma _{\sf th}}}}{{\eta \theta }}} \right)}^{N + 1 - s}}\times \\{{\left( {\frac{{{d_1^{\tau}d_2^{\tau}\gamma _{\sf th}}}}{{\eta \theta {\rho _1}}}} \right)}^{\frac{{m + s - N - 1}}{2}}} {K_{m + s - N - 1}}\left( {2\sqrt {\frac{{{d_1^{\tau}d_2^{\tau}\gamma _{\sf th}}}}{{\eta \theta {\rho _1}}}} } \right)   +\frac{2d_1^{N\tau}d_I^{N\tau}}{{\rho _1^N\rho _I^N}}\sum\limits_{s = 1}^N {\frac{{\prod\nolimits_{j = 1}^{s - 1} {\left( {1 - N - j} \right)} }}{{\left( {N - s} \right)!\left( {s - 1} \right)!}}}  \times \\{{\left( {\frac{d_1^{\tau}}{{{\rho _1}}} - \frac{d_I^{\tau}}{{{\rho _I}}}} \right)}^{1 - N - s}}
\sum\limits_{m = 0}^{N - 1} {\frac{1}{{m!}}{{\left( {\frac{{d_2^{\tau}\gamma _{\sf th}}}{{\eta \theta }}} \right)}^{N + 1 - s}}{{\left( {\frac{{{d_2^{\tau}d_I^{\tau}\gamma _{\sf th}}}}{{\eta \theta {\rho _I}}}} \right)}^{\frac{{m + s - N - 1}}{2}}}{K_{m + s - N - 1}}\left( {2\sqrt {\frac{{{d_2^{\tau}d_I^{\tau}\gamma _{\sf th}}}}{{\eta \theta {\rho _I}}}} } \right)}.
\end{multline}
\hrule
\end{figure*}

\proof See Appendix \ref{appendix:theorem:4}. \endproof
\end{theorem}

While Theorem \ref{theorem:4} is useful for the evaluation of the system's outage probability, the expression is too complex to yield much useful insights. Motivated by this, we now look into the high SNR region, and derive a simple approximation for the outage probability, which enables the characterization of the achievable diversity order of the system.

\begin{theorem}\label{theorem:5}
In the high SNR region, i.e., ${\rho _1} \to \infty $, the outage probability of the MRC/MRT scheme can be approximated as\footnote{It is worth pointing out that the result in Theorem \ref{theorem:5} holds for all cases whether the signal power and the CCI power is equal or not.}
\begin{align}\label{EOP:6}
&{P_{{I\sf{out}}}^{\sf MRC}}  \approx {\left( {\frac{{d_1^{\tau}\gamma _{{\sf{th}}}}}{{{\rho _1}}}} \right)^N}\left( {{{\left( {\frac{1}{{1 - \theta }}} \right)}^N}\sum\limits_{n = 0}^N {\frac{{{{\left( {\left( {1 - \theta } \right){\rho _I}} \right)}^n}}}{d_I^{n\tau}(N - n)!}} } \right. + \notag\\
&\left. {\frac{d_2^{N\tau}\sum\limits_{i = 0}^{N - 1} {{N-1 \choose i}{{\left( { - 1} \right)}^{N - i - 1}}\frac{{{}_2{F_1}\left( {N,2N - i - 1;2N - i;1 - \frac{{{d_1^{\tau}\rho _I}}}{{{d_I^{\tau}\rho _1}}}} \right)}}{{2N - i - 1}}}}{(\eta \theta)^N{\Gamma \left( {N + 1} \right)\Gamma \left( N \right)}} } \right).
\end{align}

\proof See Appendix \ref{appendix:theorem:5}. \endproof
\end{theorem}
For the special case where the relay is equipped with a single antenna, i.e., $N=1$, with the help of \cite[Eq. (9.121.6)]{Tables}, (\ref{EOP:6}) reduces to
\begin{align}\label{EOP:7}
{P_{{I\sf{out}}}^{\sf MRC}}
 &\approx \left( {\frac{1}{{1 - \theta }} + \frac{\rho _I}{d_I^{\tau}} + \frac{{d_2^{\tau}(\ln {\frac{\rho _1}{d_1^{\tau}}} - \ln {\frac{\rho _I}{d_I^{\tau}}})}}{{\eta \theta }}} \right)\frac{{{d_1^{\tau}\gamma _{{\sf{th}}}}}}{{{\rho _1}}}.
\end{align}
%

Theorem \ref{theorem:5} indicates that a full diversity order of $N$ is still achievable in the presence of CCI for the MRC/MRT scheme. Moreover, from (\ref{EOP:7}), we see that the effect of CCI could be either beneficial or detrimental, {depending on the relationship between $\rho_I$, $d_I^{\tau}$, $d_2^{\tau}$ $\eta$ and $\theta$, i.e., when $\frac{\rho _I}{d_I^{\tau}} - \frac{d_2^{\tau}{(\ln {\rho _I}-\ln{d_I^{\tau}})}}{{\eta \theta }}$ is positive, the CCI is detrimental, while when $\frac{\rho _I}{d_I^{\tau}} - \frac{d_2^{\tau}{(\ln {\rho _I}-\ln{d_I^{\tau}})}}{{\eta \theta }}$ is negative, the CCI becomes beneficial}, which suggests that, in wireless powered relaying systems, CCI could be potentially exploited to improve the performance.


\subsubsection{Ergodic Capacity}
Utilizing similar techniques as in Section \ref{sec:2}, we establish the following ergodic capacity upper bound:
\begin{theorem}\label{theorem:6}
If ${\rho _1} \ne {\rho _I}$,
the ergodic capacity of the MRC/MRT scheme is upper bounded by
\begin{multline}\label{EEC:1}
{C_{I\sf up}^{\sf MRC}} = {C_{{\gamma _{I1}^{\sf MRC}}}} + {C_{{\gamma _{I2}^{\sf MRC}}}} - \\
\frac{1}{2}{\log _2}\left( {1 + {e^{{\mathop{\rm E}\nolimits} \left( {\ln {\gamma _{I1}^{\sf MRC}}} \right)}} + {e^{{\mathop{\rm E}\nolimits} \left( {\ln {\gamma _{I2}^{\sf MRC}}} \right)}}} \right),
\end{multline}
where ${C_{{\gamma _{I1}^{\sf MRC}}}}$, ${C_{{\gamma _{I2}^{\sf MRC}}}}$, ${\mathop{\rm E}\nolimits} \left( {\ln {\gamma _{I1}^{\sf MRC}}} \right)$ and ${\mathop{\rm E}\nolimits} \left( {\ln {\gamma _{I2}^{\sf MRC}}} \right)$ are given by (\ref{EEC:1.1}) - (\ref{EEC:3}) shown on the next page.
\begin{figure*}
\begin{align}\label{EEC:1.1}
{C_{{\gamma _{I1}^{\sf MRC}}}} = \frac{{\left( {1 - \theta } \right){\rho _1}}}{{2d_1^{\tau}\ln 2}}\sum\limits_{m = 0}^{N - 1} {\sum\limits_{n = 0}^m {\frac{{{{\left( {\left( {1 - \theta } \right){\rho _I}} \right)}^n}}}{d_I^{n\tau}{\left( {m - n} \right)!n!}}}}
{\mathop{\rm G}\nolimits} _{1,[1:1],0,[1:1]}^{1,1,1,1,1}\left(^\frac{\left( {1 - \theta } \right){\rho _1}}{d_1^{\tau}}_\frac{\left( {1 - \theta } \right){\rho _I}}{d_I^{\tau}} \middle| \substack{{m + 1}\\
{0; - n}\\
 - \\
{0;0}}
\right),
\end{align}
\hrule
\end{figure*}
%
%
\begin{figure*}
\begin{multline}\label{EEC:1.2}
{C_{{\gamma _{I2}^{\sf MRC}}}} = \frac{d_1^{N\tau}d_2^{N\tau}}{{\rho _1^N\rho _I^N2\ln 2}}\sum\limits_{s = 1}^N {\frac{{\prod\nolimits_{j = 1}^{s - 1} {\left( {1 - N - j} \right)} }}{{\left( {N - s} \right)!\left( {s - 1} \right)!}}{{\left( {\frac{d_I^{\tau}}{{{\rho _I}}} - \frac{d_1^{\tau}}{{{\rho _1}}}} \right)}^{1 - N - s}}  }
\sum\limits_{m = 0}^{N - 1} \frac{1}{{m!}}{{\left( {\frac{d_2^{\tau}}{{\eta \theta }}} \right)}^{N + 1 - s}}\times\\
G_{1,3}^{3,1}\left( \frac{d_1^{\tau}d_2^{\tau}}{{\eta \theta {\rho _1}}}\middle|^
{s - N - 1}_
{s - N - 1,m + s - N - 1,0}
\right)  +
\frac{d_1^{N\tau}d_2^{N\tau}}{{\rho _1^N\rho _I^N2\ln 2}}\sum\limits_{s = 1}^N {\frac{{\prod\nolimits_{j = 1}^{s - 1} {\left( {1 - N - j} \right)} }}{{\left( {N - s} \right)!\left( {s - 1} \right)!}}{{\left( {\frac{d_1^{\tau}}{{{\rho _1}}} - \frac{d_I^{\tau}}{{{\rho _I}}}} \right)}^{1 - N - s}}} \\
 \times \sum\limits_{m = 0}^{N - 1} {\frac{1}{{m!}}{{\left( {\frac{d_2^{\tau}}{{\eta \theta }}} \right)}^{N + 1 - s}}G_{1,3}^{3,1}\left( \frac{d_2^{\tau}d_I^{\tau}}{{\eta \theta {\rho _I}}}\middle|^
{s - N - 1}_
{s - N - 1,m + s - N - 1,0} \right)},
\end{multline}
\hrule
\end{figure*}
\begin{figure*}
\begin{multline}\label{EEC:2}
{\mathop{\rm E}\nolimits} \left( {\ln {\gamma _{I1}^{\sf MRC}}} \right) = \ln \left( {\left( {1 - \theta } \right){\rho _1}} \right)-\ln{d_1^{\tau}} + \psi \left( 1 \right) - {e^{\frac{d_I^{\tau}}{{\left( {1 - \theta } \right){\rho _I}}}}}G_{2,3}^{3,0}\left( \frac{d_I^{\tau}}{{\left( {1 - \theta } \right){\rho _I}}}\middle|^
{1,1}_
{0,0,1}\right) + \\
\sum\limits_{m = 1}^{N - 1} {\sum\limits_{n = 0}^m {\frac{{{{\left( {\left( {1 - \theta } \right){\rho _I}} \right)}^{n - m}}}}{{(m - n)!d_I^{(n-m)\tau}}}\Gamma \left( m \right)\Psi \left( {m,m - n;\frac{d_I^{\tau}}{{\left( {1 - \theta } \right){\rho _I}}}} \right)} },
\end{multline}
\hrule
\end{figure*}
\begin{figure*}
\begin{multline}\label{EEC:3}
{\mathop{\rm E}\nolimits} \left( {\ln {\gamma _{I2}^{\sf MRC}}} \right) = \ln \eta \theta -\ln {d_2^{\tau}} + \psi \left( N \right) +
\frac{d_I^{N\tau}}{{\rho _I^N}}\sum\limits_{s = 1}^N \frac{{\prod\nolimits_{j = 1}^{s - 1} {\left( {1 - N - j} \right)} }}{{\left( {s - 1} \right)!}}{{\left( {\frac{d_I^{\tau}}{{{\rho _I}}} - \frac{d_1^{\tau}}{{{\rho _1}}}} \right)}^{1 - N - s}}\left({\frac{\rho _1}{d_1^{\tau}}}\right)^{1 - s}\times\\ \left( {\psi \left( {N - s + 1} \right) + \ln {{\rho _1}}}-\ln{d_1^{\tau}} \right)
 + \frac{d_1^{N\tau}}{{\rho _1^N}}\sum\limits_{s = 1}^N {\frac{{\prod\nolimits_{j = 1}^{s - 1} {\left( {1 - N - j} \right)} }}{{\left( {s - 1} \right)!}}{{\left( {\frac{d_1^{\tau}}{{{\rho _1}}} - \frac{d_I^{\tau}}{{{\rho _I}}}} \right)}^{1 - N - s}}\left(\frac{\rho _I}{d_I^{\tau}}\right)^{1 - s}}\times\\\left( {\psi \left( {N - s + 1} \right) + \ln {\rho _I}-\ln {d_I^{\tau}}} \right).
\end{multline}
\hrule
\end{figure*}

\proof See Appendix \ref{appendix:theorem:6}. \endproof
\end{theorem}


\subsubsection{Optimal $\theta$ Analysis}\label{sec:4.3}
We now study the optimal value of $\theta$ minimizing the outage probability. Based on the high SNR approximation for ${P_{I\sf out}^{\sf MRC}}$ in (\ref{EOP:6}), the optimal $\theta$ can be found as:
\begin{proposition}\label{prop:2}
The optimal $\theta$ is a root of the following polynomial
\begin{align}\label{EOT:2}
\sum\limits_{n = 0}^{N - 1} {{\cal A}\left( n \right){{\left( {1 - \theta } \right)}^{n - N - 1}}}  - \frac{\cal B}{{{\theta ^{N + 1}}}} = 0,
\end{align}
where ${\cal A}\left( n \right) = \frac{\frac{\rho _I^n}{d_I^{n\tau}}}{{\left( {N - n - 1} \right)!}}$, ${\cal B} = \frac{{d_2^{N\tau}}}{{{\eta ^N}{\Gamma ^2}\left( N \right)}}\sum\limits_{i = 0}^{N - 1} {{N-1\choose i}{{\left( { - 1} \right)}^{N - i - 1}}\frac{{{}_2{F_1}\left( {N,2N - i - 1;2N - i;1 - \frac{{{\rho _I}{d_1^{\tau}}}}{{{\rho _1}{d_I^{\tau}}}}} \right)}}{{2N - i - 1}}}$ and $0<\theta<1$.

\proof
{The result is derived by following the same steps as in the proof of Proposition \ref{prop:1}.}
\endproof
\end{proposition}

In the special case of $N=1$, the optimal solution can be given in closed-form as follows:
\begin{align}\label{EOT:3}
{\theta_{\sf MRC}^{\sf opt}} = \frac{{\sqrt {\frac{{{d_I^{\tau}\rho _1}\left( {\ln {\rho _I} - \ln {\rho _1}-\ln{d_I^{\tau}}+\ln{d_1^{\tau}}} \right)}}{{\eta \left( {{d_1^{\tau}\rho _I} - {d_I^{\tau}\rho _1}} \right)}}} }}{{1 + {\sqrt {\frac{{{d_I^{\tau}\rho _1}\left( {\ln {\rho _I} - \ln {\rho _1}-\ln{d_I^{\tau}}+\ln{d_1^{\tau}}} \right)}}{{\eta \left( {{d_1^{\tau}\rho _I} - {d_I^{\tau}\rho _1}} \right)}}} } }}.
\end{align}
This simple expression is quite informative, and it can be observed that the optimal $\theta$ in (\ref{EOT:3}) is a decreasing function of $\eta$ and $\rho_I$, and an increasing function of $\rho_1$, which can be explained as follows:
\begin{itemize}
  \item As $\eta$ increases, more transmission power can be collected at the relay, hence the bottleneck of the system performance lies in the SINR of the signal at the input of the information receiver. As a result, we should choose a smaller $\theta$ to improve the first hop performance.
  \item A large $\rho_I$ provides more energy, while at the same time reduces the SINR of the first hop transmission. Hence, a smaller $\theta$ should be chosen to compensate the loss of the SINR.
  \item For large $\rho_1$, in general the first hop transmission quality is quite good, hence, it is beneficial to have more energy at the relay, i.e., a larger $\theta$ is desirable.
\end{itemize}

\subsection{ZF/MRT Scheme}
For the ZF/MRT scheme, the relay utilizes the available multiple antennas to completely eliminate the CCI. To ensure this is possible, the number of the antennas equipped at the relay should be greater than the number of interferers. Hence, for the ZF/MRT scheme, it is assumed that $N > 1$. According to \cite{G.Zhu}, the optimal combining vector ${\bf{w}}_1$ is given by
$
{\bf{w}}_1 = \frac{{{\bf{h}}_1^\dag {\bf{P}}}}{{\sqrt {{\bf{h}}_1^\dag {\bf{P}}{{\bf{h}}_1}} }},
$
where ${\bf{P}} = {{\bf{I}}_N} - {{\bf{h}}_I}{\left( {{\bf{h}}_I^\dag {{\bf{h}}_I}} \right)^{ - 1}}{\bf{h}}_I^\dag $.
Therefore, the end-to-end SINR of the ZF/MRT scheme can be expressed as
\begin{align}\label{ZF:1}
{\gamma _I^{\sf ZF}} = \frac{{{\gamma _{I1}^{\sf ZF}}{\gamma _{I2}^{\sf ZF}}}}{{{\gamma _{I1}^{\sf ZF}} + {\gamma _{I2}^{\sf ZF}} + 1}},
\end{align}
where ${\gamma _{I1}^{\sf ZF}} = \left| {{\bf{h}}_1^\dag {\bf{P}}{{\bf{h}}_1}} \right|\frac{(1 - \theta){\rho _1}}{d_1^{\tau}}$, ${\gamma _{I2}^{\sf ZF}} = \frac{\eta \theta}{d_2^{\tau}} \left( {\frac{\rho _1}{d_1^{\tau}}\left\| {{{\bf{h}}_1}} \right\|_F^2 + \frac{\rho _I}{d_I^{\tau}}\left\| {{{\bf{h}}_I}} \right\|_F^2} \right)\left\| {{{\bf{h}}_2}} \right\|_F^2$.

\subsubsection{Outage Probability}
We first present the following outage lower bound:
%
%

\begin{theorem}\label{theorem:7}
If ${\rho _1} \ne {\rho _I}$, the outage probability of the ZF/MRT scheme can be lower bounded as
\begin{align}\label{ZFOP:3}
{P_{{I\sf{out}}}^{\sf LZF}} = 1 - {F_1^{\sf ZF}}{F_2^{\sf ZF}},
\end{align}
where ${F_1^{\sf ZF}} = \frac{{\Gamma \left( {N - 1,\frac{{{d_1^{\tau}\gamma _{{\rm{th}}}}}}{{\left( {1 - \theta } \right){\rho _1}}}} \right)}}{{\Gamma \left( {N - 1} \right)}}$ and ${F_2^{\sf ZF}} = {F_2^{\sf MRC}}$.

\proof According to \cite{G.Zhu}, the c.d.f. of ${\gamma _{I1}^{\sf ZF}}$ is given by
\begin{align}\label{PZFOP:1}
{F_{{\gamma _{I1}^{\sf ZF}}}}\left( x \right) = 1 - \frac{{\Gamma \left( {N - 1,\frac{d_1^{\tau}x}{{\left( {1 - \theta } \right){\rho _1}}}} \right)}}{{\Gamma \left( {N - 1} \right)}}.
\end{align}
Then, the desired result can be obtained by following the similar lines as in the proof of Theorem \ref{theorem:4}.
\endproof
\end{theorem}


To gain further insights, we now look into the high SNR region, and present a simple and informative approximation for the outage probability.
\begin{theorem}\label{theorem:8}
In the high SNR region, i.e., ${\rho _1} \to \infty $, the outage probability of the ZF/MRT scheme can be approximated as
\begin{align}\label{ZFOP:6}
{P_{{I\sf{out}}}^{\sf ZF}}  \approx \frac{1}{{\left( {N - 1} \right)!}}{\left( {\frac{{{d_1^{\tau}\gamma _{{\rm{th}}}}}}{{\left( {1 - \theta } \right){\rho _1}}}} \right)^{N - 1}}.
\end{align}

\proof
With the help of the asymptotic expansion of incomplete gamma function given in \cite[Eq. (8.354.2)]{Tables}, it is easy to note that the c.d.f. of ${\gamma _{I1}^{\sf ZF}}$ can be approximated as
\begin{align}\label{PZFOP:2}
{F_{{\gamma _{I1}^{\sf ZF}}}}\left( x \right) \approx \frac{1}{{\left( {N - 1} \right)!}}{\left( {\frac{d_1^{\tau}x}{{\left( {1 - \theta } \right){\rho _1}}}} \right)^{N - 1}}.
\end{align}
Then, utilizing (\ref{PZFOP:2}) and following the similar lines as in the proof of Theorem \ref{theorem:5}, we can obtain
\begin{multline}\label{PZFOP:3}
{P_{{I\sf{out}}}^{\sf ZF}}  \approx \frac{1}{{\left( {N - 1} \right)!}}{\left( {\frac{{d_1^{\tau}{\gamma _{{\rm{th}}}}}}{{\left( {1 - \theta } \right){\rho _1}}}} \right)^{N - 1}} + \\ \frac{{\left( {\frac{{{d_1^{\tau}d_2^{\tau}\gamma _{{\rm{th}}}}}}{{\eta \theta {\rho _1}}}} \right)^N}}{{\Gamma \left( {N + 1} \right)\Gamma \left( N \right)}}
\sum\limits_{i = 0}^{N - 1} {{N-1\choose i}}{{\left( { - 1} \right)}^{N - i - 1}}\times \\
\frac{{{}_2{F_1}\left( {N,2N - i - 1;2N - i;1 - \frac{{{d_1^{\tau}\rho _I}}}{{{d_I^{\tau}\rho _1}}}} \right)}}{{2N - i - 1}}.
\end{multline}
The desired result follows by noticing that the second term is negligible compared with the first term in (\ref{PZFOP:3}).
\endproof
\end{theorem}

Theorem \ref{theorem:8} indicates that the achievable diversity order of the ZF/MRT scheme is $N-1$. Compared with the MRC/MRT scheme, the ZF/MRT scheme incurs a diversity loss of one. This is an intuitive and satisfying result since one degree of freedom is used for the elimination of the CCI.

\subsubsection{Ergodic Capacity}
We now look into the ergodic capacity of the system, and we can establish the following upper bound of the ergodic capacity:
\begin{theorem}\label{theorem:9}
If ${\rho _1} \ne {\rho _I}$,
the ergodic capacity of the ZF/MRT scheme is upper bounded by
\begin{align}\label{ZFEC:1}
{C_{I\sf up}^{\sf ZF}} = {C_{{\gamma _{I1}^{\sf ZF}}}} + {C_{{\gamma _{I2}^{\sf ZF}}}} - \frac{1}{2}{\log _2}\left( {1 + {e^{{\mathop{\rm E}\nolimits} \left( {\ln {\gamma _{I1}^{\sf ZF}}} \right)}} + {e^{{\mathop{\rm E}\nolimits} \left( {\ln {\gamma _{I2}^{\sf ZF}}} \right)}}} \right),
\end{align}

where ${C_{{\gamma _{I1}^{\sf ZF}}}} = \frac{e^{\frac{d_1^{\tau}}{{\left( {1 - \theta } \right){\rho _1}}}}}{{2\ln 2}}\sum\limits_{k = 0}^{N - 2} {{{\left( {\frac{d_1^{\tau}}{{\left( {1 - \theta } \right){\rho _1}}}} \right)}^k}\Gamma \left( { - k,\frac{d_1^{\tau}}{{\left( {1 - \theta } \right){\rho _1}}}} \right)}$, ${\mathop{\rm E}\nolimits} \left( {\ln {\gamma _{I1}^{\sf ZF}}} \right) = \ln \left( {\left( {1 - \theta } \right){\rho _1}} \right) -\ln {d_1^{\tau}}+ \psi \left( {N - 1} \right)$, ${C_{{\gamma _{I2}^{\sf ZF}}}} = {C_{{\gamma _{I2}^{\sf MRC}}}}$ and ${\mathop{\rm E}\nolimits} \left( {\ln {\gamma _{I2}^{\sf ZF}}} \right) = {\mathop{\rm E}\nolimits} \left( {\ln {\gamma _{I2}^{\sf MRC}}} \right)$.

\proof With the help of the c.d.f. of ${\gamma _{I1}^{\sf ZF}}$ given in (\ref{PZFOP:1}) and following the similar lines as in the proof of Theorem \ref{theorem:6} yields the desired result.
\endproof
\end{theorem}


\subsubsection{Optimal $\theta$ Analysis}
We now study the optimal $\theta$ minimizing the outage probability. Based on the high SNR approximation for ${P_{I\sf out}^{\sf ZF}}$ in (\ref{PZFOP:3}), the optimal $\theta$ can be found as:

\begin{proposition}\label{prop:3}
The optimal $\theta$ is a root of the following polynomial
\begin{align}\label{ZFOT:2}
\frac{{\cal A}_1}{{{(1-\theta) ^{N}}}}  - \frac{{\cal B}_1}{{{\theta ^{N + 1}}}} = 0,
\end{align}
where ${\cal A}_1 = \frac{{1}}{{\left( {N - 2} \right)!}}$, ${\cal B}_1 = \frac{d_2^{N\tau}d_1^{\tau}\gamma_{\sf th}}{{{\eta ^N}{\Gamma ^2}\left( N \right)}\rho_1}\sum\limits_{i = 0}^{N - 1} {{N-1\choose i}{{\left( { - 1} \right)}^{N - i - 1}}\frac{{{}_2{F_1}\left( {N,2N - i - 1;2N - i;1 - \frac{{{d_1^{\tau}\rho _I}}}{{{d_I^{\tau}\rho _1}}}} \right)}}{{2N - i - 1}}}$ and $0<\theta<1$.

\proof
{The result is derived by following the same steps as in the proof of Proposition \ref{prop:1}.}
\endproof
\end{proposition}

\subsection{MMSE/MRT Scheme}
The ZF scheme completely eliminates the CCI at the relay, which however may cause an elevated noise level. In contrast, the MMSE scheme does not fully eliminate the CCI, instead, it provides the optimum trade-off between interference suppression and noise enhancement.
According to \cite{G.Zhu}, ${\bf w}_1$ is given by
\begin{align}
{\bf{w}}_1 = {\bf{h}}_1^\dag {\left( {{{\bf{h}}_1}{\bf{h}}_1^\dag  + {{\bf{h}}_I}{\bf{h}}_I^\dag  + \frac{d_I^{\tau}}{{\left( {1 - \theta } \right){\rho _I}}}{\bf{I}}} \right)^{ - 1}}.
\end{align}
Therefore, the end-to-end SINR of the MMSE/MRT scheme can be expressed as
\begin{align}\label{MMSE:1}
{\gamma _I^{\sf MMSE}} = \frac{{{\gamma _{I1}^{\sf MMSE}}{\gamma _{I2}^{\sf MMSE}}}}{{{\gamma _{I1}^{\sf MMSE}} + {\gamma _{I2}^{\sf MMSE}} + 1}},
\end{align}
where ${\gamma _{I1}^{\sf MMSE}} = \frac{{{d_I^{\tau}\rho _1}}}{{{d_1^{\tau}\rho _I}}}{\bf{h}}_1^\dag {{\bf{R}}^{ - 1}}{{\bf{h}}_1}$, ${\bf{R}} = {{\bf{h}}_I}{\bf{h}}_I^\dag  + \frac{d_I^{\tau}}{{\left( {1 - \theta } \right){\rho _I}}}{\bf{I}}$ and ${\gamma _{I2}^{\sf MMSE}} = \frac{\eta \theta}{d_2^{\tau}} \left( {\frac{\rho _1}{d_1^{\tau}}\left\| {{{\bf{h}}_1}} \right\|_F^2 + \frac{\rho _I}{d_I^{\tau}}\left\| {{{\bf{h}}_I}} \right\|_F^2} \right)\left\| {{{\bf{h}}_2}} \right\|_F^2$.

\subsubsection{Outage Probability}
%
%

\begin{theorem}\label{theorem:10}
If ${\rho _1} \ne {\rho _I}$, the outage probability of the MMSE/MRT scheme can be lower bounded as
\begin{align}\label{MMSEOP:3}
{P_{{I\sf{out}}}^{\sf LMMSE}} = 1 - {F_1^{\sf MMSE}}{F_2^{\sf MMSE}},
\end{align}
where ${F_1^{\sf MMSE}} = \frac{{\Gamma \left( {N,\frac{{{d_1^{\tau}\gamma _{{\sf{th}}}}}}{{\left( {1 - \theta } \right){\rho _1}}}} \right)}}{{\Gamma \left( N \right)}} - \frac{{{e^{ - \frac{{{d_1^{\tau}\gamma _{{\sf{th}}}}}}{{\left( {1 - \theta } \right){\rho _1}}}}}\left( {1 - \theta } \right){\rho _I}}}{{d_I^{\tau}\Gamma( N)}}{\left( {\frac{{{d_1^{\tau}\gamma _{{\sf{th}}}}}}{{\left( {1 - \theta } \right){\rho _1}}}} \right)^N}{}_2{F_1}\left( {2,1;2; - \frac{{{d_1^{\tau}\rho _I}}}{{{d_I^{\tau}\rho _1}}}{\gamma _{{\sf{th}}}}} \right)$
and ${F_2^{\sf MMSE}} = {F_2^{\sf MRC}}$.

\proof According to \cite{G.Zhu} we know that the c.d.f. of ${\gamma _{I1}^{\sf MMSE}}$ is given by
\begin{multline}\label{PMMSEOP:1}
{F_{{\gamma _{I1}^{\sf MMSE}}}}\left( x \right) = 1 - \frac{{\Gamma \left( {N,\frac{d_1^{\tau}x}{{\left( {1 - \theta } \right){\rho _1}}}} \right)}}{{\Gamma \left( N \right)}} + \frac{{{e^{ - \frac{d_1^{\tau}x}{{\left( {1 - \theta } \right){\rho _1}}}}}\left( {1 - \theta } \right){\rho _I}}}{{d_I^{\tau}\Gamma \left( N \right)}}\\ \times {\left( {\frac{d_1^{\tau}x}{{\left( {1 - \theta } \right){\rho _1}}}} \right)^N}{}_2{F_1}\left( {2,1;2; - \frac{{{d_1^{\tau}\rho _I}}}{{{d_I^{\tau}\rho _1}}}x} \right).
\end{multline}
Then, following the similar lines as in the proof of Theorem \ref{theorem:4}, we can obtain the desired result.
\endproof
\end{theorem}


To gain further insights, we now look into the high SNR region, and present a simple approximation for the outage probability.

\begin{theorem}\label{theorem:11}
In the high SNR region, i.e., ${\rho _1} \to \infty $, the outage probability of the MMSE/MRT scheme can be approximated as
\begin{multline}\label{MMSEOP:6}
{P_{{I\sf{out}}}^{\sf MMSE}}  \approx {\left( {\frac{{{d_1^{\tau}\gamma _{{\sf{th}}}}}}{{{\rho _1}}}} \right)^N}\left( {\left( {\frac{1}{{N!}} + \frac{{\left( {1 - \theta } \right){\rho _I}}}{{d_I^{\tau}\Gamma \left( N \right)}}} \right)\frac{1}{{{{\left( {1 - \theta } \right)}^N}}} } \right.+ \\
\frac{{{{\left( {\frac{d_2^{\tau}}{{\eta \theta }}} \right)}^N}}}{{\Gamma \left( {N + 1} \right)\Gamma \left( N \right)}}
\sum\limits_{i = 0}^{N - 1} {{N-1\choose i}}{{\left( { - 1} \right)}^{N - i - 1}}\times\\
\left. {\frac{{{}_2{F_1}\left( {N,2N - i - 1;2N - i;1 - \frac{{{d_1^{\tau}\rho _I}}}{{{d_I^{\tau}\rho _1}}}} \right)}}{{2N - i - 1}} } \right).
\end{multline}

\proof
After some simple manipulation the c.d.f. of ${\gamma _{I1}^{\sf MMSE}}$ can be approximated as
\begin{align}\label{PMMSEOP:2}
{F_{{\gamma _{I1}^{\sf MMSE}}}}\left( x \right) \approx \left( {\frac{1}{{N!}} + \frac{{\left( {1 - \theta } \right){\rho _I}}}{{d_I^{\tau}\Gamma \left( N \right)}}} \right){\left( {\frac{{{d_1^{\tau}\gamma _{{\rm{th}}}}}}{{\left( {1 - \theta } \right){\rho _1}}}} \right)^N}.
\end{align}
Then, following the similar lines as in the proof of Theorem \ref{theorem:5}, we can obtain the desired result.
\endproof
\end{theorem}

Theorem \ref{theorem:11} indicates that the MMSE/MRT scheme achieves a diversity order of $N$, the same as the MRC/MRT scheme.

{A close observation of (\ref{EOP:6}), (\ref{PZFOP:3}) and (\ref{MMSEOP:6}) reveals that the difference among all three schemes only lies in their first terms, which can be expressed as follows:}
{\begin{align}
&{a_{\sf MRC}}= {\sum\limits_{n = 0}^N {\frac{{{{\left( {\left( {1 - \theta } \right){\rho _I}} \right)}^n}}}{{d_I^{n\tau }(N - n)!}}} },\notag\\
&a_{\sf ZF}=\frac{1}{(N-1)!},\notag\\
&a_{\sf MMSE}={\frac{1}{N!}+\frac{(1-\theta)\rho_I}{d_I^\tau(N-1)!}}.
\end{align}
}

{It can be easily observed that $a_{\sf MMSE}$ is strictly smaller than $a_{\sf MRC}$, since $a_{\sf MMSE}$ only includes the first two terms of $a_{\sf MRC}$. As such, we conclude that the MMSE/MRT scheme always achieves a strictly better outage performance than the MRC/MRT scheme due to the higher array gain. For the ZF/MRT scheme, although a diversity loss leads to its inferior performance in the high SNR region, it should be noted that $a_{\sf ZF}$ is generally smaller than $a_{\sf MRC}$, which means that the ZF/MRC scheme has a larger array gain than the MRC/MRT scheme. Therefore, in the low SNR region, the ZF/MRT scheme may achieve better outage performance than the MRC/MRT scheme.}

\subsubsection{Ergodic Capacity}
%
\begin{theorem}\label{theorem:12}
If ${\rho _1} \ne {\rho _I}$,
the ergodic capacity of the MMSE/MRT scheme is upper bounded by
\begin{multline}\label{MMSEEC:1}
{C_{I\sf up}^{\sf MMSE}} = {C_{{\gamma _{I1}^{\sf MMSE}}}} + {C_{{\gamma _{I2}^{\sf MMSE}}}} - \\
\frac{1}{2}{\log _2}\left( {1 + {e^{{\mathop{\rm E}\nolimits} \left( {\ln {\gamma _{I1}^{\sf MMSE}}} \right)}} + {e^{{\mathop{\rm E}\nolimits} \left( {\ln {\gamma _{I2}^{\sf MMSE}}} \right)}}} \right),
\end{multline}
where
\begin{multline}
{C_{{\gamma _{I1}^{\sf MMSE}}}} =\frac{{e^{\frac{d_1^{\tau}}{{\left( {1 - \theta } \right){\rho _1}}}}}}{{2\ln 2}}
\sum\limits_{k = 0}^{N - 1} {{{\frac{d_1^{k\tau}\Gamma \left( { - k,\frac{d_1^{\tau}}{{\left( {1 - \theta } \right){\rho _1}}}} \right)}{({\left( {1 - \theta } \right){\rho _1}})^k}}}}  - \\ \frac{{{{\left( {1 - \theta } \right)}^3}\rho _I^2{\rho _1}}}{{2\ln 2\;\Gamma \left( N \right)d_I^{2\tau}d_1^{\tau}}}{\mathop{\rm G}\nolimits} _{1,[1:2],0,[1:2]}^{1,1,2,1,1}\left(^{\frac{(1 - \theta){\rho _1}}{d_1^{\tau}}}_{\frac{(1 - \theta){\rho _I}}{d_I^{\tau}}} \middle| \substack{{N + 2}\\
{0;\left( { - 2, - 1} \right)}\\
 - \\
{0;\left( { - 1, - 2} \right)}}
\right),
\end{multline}
\begin{multline}
{\mathop{\rm E}\nolimits} \left( {\ln {\gamma _{I1}^{\sf MMSE}}} \right) = \psi \left( N \right) + \ln \left( {(1 - \theta ){\rho _1}} \right) - \ln{d_1^{\tau}}-\\
 \frac{{{{\left( {\left( {1 - \theta } \right){\rho _I}} \right)}^2}}}{{d_I^{2\tau}\Gamma ( N )}}{\mathop{\rm G}\nolimits} _{3,2}^{1,3}\left(\frac{(1 - \theta){\rho _I}}{d_I^{\tau}}\middle|^
{ - N, - 2, - 1}_
{ - 1, - 2} \right),
\end{multline}
and ${C_{{\gamma _{I2}^{\sf MMSE}}}} = {C_{{\gamma _{I2}^{\sf MRC}}}}$ as well as ${\mathop{\rm E}\nolimits} \left( {\ln {\gamma _{I2}^{\sf MMSE}}} \right) = {\mathop{\rm E}\nolimits} \left( {\ln {\gamma _{I2}^{\sf MRC}}} \right)$.

\proof With the help of the c.d.f. of ${\gamma _{I1}^{\sf MMSE}}$ given in (\ref{PMMSEOP:1}) and follows the similar lines as in the proof of Theorem \ref{theorem:6} yields the desired result.
\endproof
\end{theorem}


\subsubsection{Optimal $\theta$ Analysis}\label{sec:4.4}
We now study the optimal $\theta$ minimizing the outage probability. Based on the high SNR approximation for ${P_{I\sf out}^{\sf MMSE}}$ in (\ref{MMSEOP:6}), the optimal $\theta$ can be found as:

\begin{proposition}\label{prop:4}
The optimal $\theta$ is a root of the following equation
\begin{align}\label{MMSEOT:2}
\frac{1}{{{(1-\theta) ^{N+1}\Gamma(N)}}} + \frac{(N-1)\rho_I}{{{d_I^{\tau}(1-\theta)^{N}\Gamma(N)}}}  - \frac{{\cal B}}{{{\theta ^{N + 1}}}} = 0,
\end{align}
where ${\cal B}$ have been defined in (\ref{EOT:2}) and $0<\theta<1$.

\proof
{The result is derived by following the same steps as in the proof of Proposition \ref{prop:1}.}
\endproof
\end{proposition}


\section{Numerical Results and Discussion}\label{sec:5}
In this section, we present numerical results to validate the analytical expressions presented in Section IV, and investigate the impact of various key system parameters on the system's performance. Unless otherwise specified, we set ${\gamma _{\sf th}} = 0\;{\mbox{dB}}$, {$\eta = 0.8$}, $\theta = 0.5$, $\rho_I = 9.5\;{\mbox{dB}}$, {$\tau=2$ and $d_1=d_2=d_I=1$.}

\subsection{Effect of Multiple Antennas}
\begin{figure}[ht]
  \centering
  \subfigure[Outage probability]{\label{fig:2a}\includegraphics[width=0.4\textwidth]{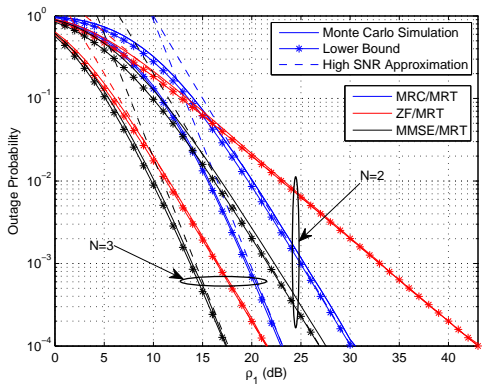}}
  \hspace{0.2in}
  \subfigure[Ergodic capacity]{\label{fig:2b}\includegraphics[width=0.4\textwidth]{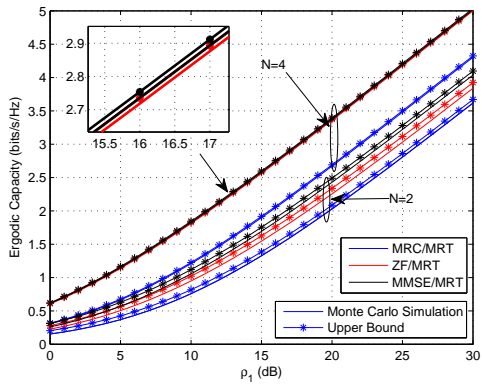}}
  \caption{Impact of $N$ on the system performance.}
  \label{fig:fig2}
\end{figure}
Fig. \ref{fig:fig2} illustrates the impact of antenna number $N$ on the outage probability and ergodic capacity. It can be readily observed from Fig. \ref{fig:2a} that for all the three considered schemes,
the proposed lower bounds {in (\ref{EOP:3}), (\ref{ZFOP:3}) and (\ref{MMSEOP:3})} are sufficiently tight across the entire SNR range of interest, especially when $N$ is large, and become almost exact in the high SNR region, while the high SNR approximations {in (\ref{EOP:6}), (\ref{ZFOP:6}) and (\ref{MMSEOP:6})} work quite well even at moderate SNR values (i.e., ${\rho _1} = 20\;\mbox{dB}$). In addition, we see that both the MRC/MRT and MMSE/MRT schemes achieve the full diversity order of $N$, while the ZF/MRT scheme only achieves a diversity order of $N-1$, which is consistent with our analytical results. Moreover, the MMSE/MRT scheme always attains the best outage performance among all three proposed schemes, and the ZF/MRT scheme outperforms the MRC/MRT scheme in the low SNR region, while the opposite holds in the high SNR region.

From Fig. \ref{fig:2b}, we see that, for all three schemes, the proposed ergodic capacity upper bounds {in (\ref{EEC:1}), (\ref{ZFEC:1}) and (\ref{MMSEEC:1})} are sufficiently tight across the entire SNR range of interest.
In addition, we observe the intuitive result that increasing $N$ results in an improvement of the ergodic
capacity. Moreover, the MMSE/MRT scheme always has
the best performance, while the ZF/MRT scheme is slightly inferior, and the performance gap between them disappears as $N$ increases. On the other hand, the MRC/MRT scheme always yields the lowest ergodic capacity, and as $N$ increases, the performance gap becomes more pronounced.

\subsection{Effect of CCI}
\begin{figure}[ht]
  \centering
  \subfigure[Outage probability]{\label{fig:3a}\includegraphics[width=0.4\textwidth]{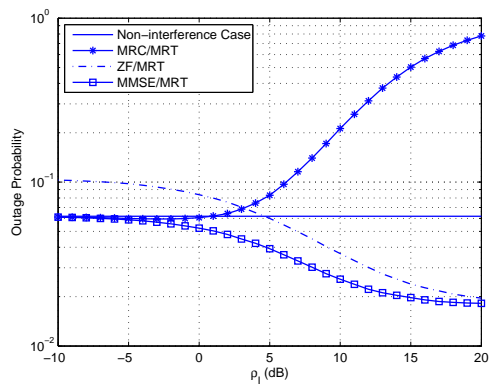}}
  \hspace{0.2in}
  \subfigure[Ergodic capacity]{\label{fig:3b}\includegraphics[width=0.4\textwidth]{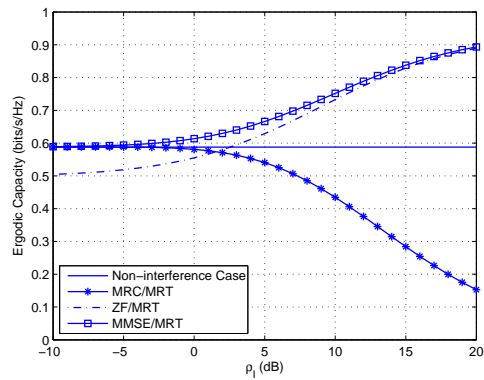}}
  \caption{Impact of CCI on the system performance.}
  \label{fig:fig3}
\end{figure}


Fig. \ref{fig:fig3} investigates the impact CCI on the system performance. The scenario without CCI is also plotted for comparison. It can be readily observed from Fig. \ref{fig:3a} that the outage probability of the MRC/MRT scheme decreases slightly for smaller $\rho_I$ (i.e., $\rho_I<0\mbox{ dB}$), and then increases as the interference becomes stronger. {This phenomenon clearly indicates that the CCI can cause either beneficial or harmful effect on the system's performance.} {This is because that CCI provides additional energy but at the same time corrupts the desired signal.}
For the MRC/MRT scheme, when the CCI is too strong, the disadvantage of the CCI becomes the dominant performance limiting factor. However, with sophisticated interference mitigation schemes, e.g., the ZF/MRT and MMSE/MRT schemes, such undesirable effect could be eliminated. As shown in these two schemes, the outage probability decreases monotonically as $\rho_I$ increases. Moreover, for the MMSE/MRT scheme, CCI is always desirable, while for the ZF/MRT scheme, whether CCI is beneficial or not depends on its power.

\subsection{Effect of the Distance}
\begin{figure}[ht]
  \centering
  \subfigure[Impact of relay location, $d_2=5-d_1$, $d_I=3$.]{\label{fig:6a}\includegraphics[width=0.4\textwidth]{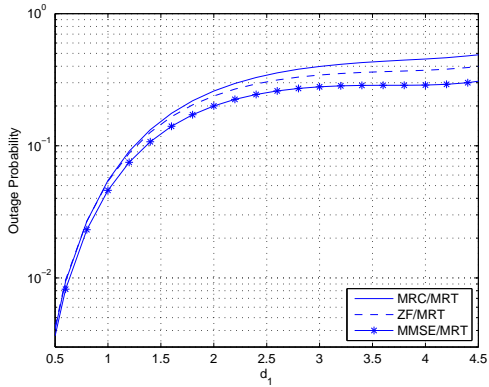}}
  \hspace{0.2in}
  \subfigure[Impact of interferer location, $d_1=2$, $d_2=3$.]{\label{fig:6b}\includegraphics[width=0.4\textwidth]{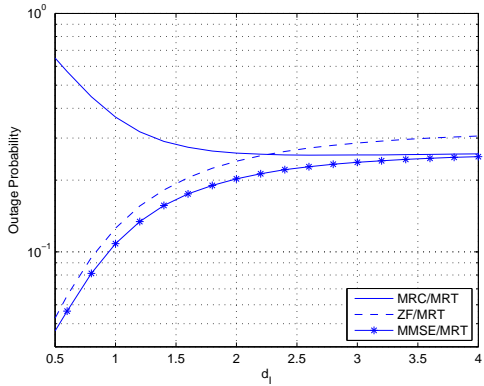}}
  \caption{Impact of distance on the system performance.}
  \label{fig:fig6}
\end{figure}
Fig. \ref{fig:fig6} shows the effect of the node distances on the system's outage probability. Unlike the conventional dual-hop system, where it is in general desirable to place the relay in the middle of the source and the destination, Fig. \ref{fig:6a} indicates that in the energy harvesting scenario, the optimal relay location tends to be close to the source. This observation implies that, the quality of the first hop channel is more important than that of the second hop channel. This is quite intuitive since the quality of the first hop channel not only affects the received signal power at the relay but also determines the available power for the second hop transmission. As shown in Fig. \ref{fig:6b}, as the distance of the CCI increases, the outage performance of ZF/MRT and MMSE/MRT schemes deteriorates. In contrast, the outage performance of the MRC/MRT scheme improves. {This is also intuitive, since increasing the distance reduces the received power at the relay, which in turn deteriorates the performance of the ZF/MRT and MMSE/MRT schemes, since strong CCI is desirable for the both the ZF/MRT and MMSE/MRT schemes as illustrated in Fig. \ref{fig:fig3}.}

\subsection{Effect of Power Splitting Ratio $\theta$}
\begin{figure}[ht]
  \centering
  \subfigure[Outage probability]{\label{fig:4a}\includegraphics[width=0.4\textwidth]{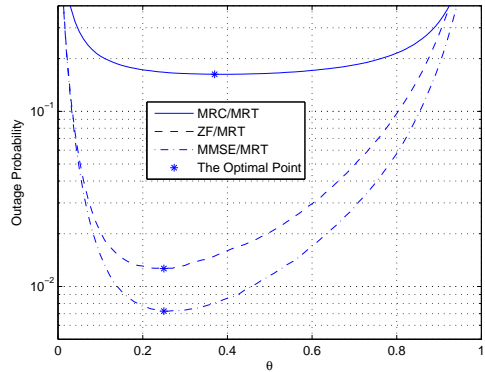}}
  \hspace{0.2in}
  \subfigure[Ergodic capacity]{\label{fig:4b}\includegraphics[width=0.4\textwidth]{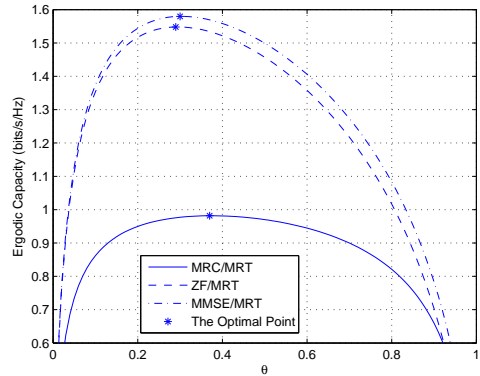}}
  \caption{The optimal power splitting ratio $\theta$.}
  \label{fig:fig4}
\end{figure}
Fig. \ref{fig:fig4} investigates the the impact of the power splitting ratio $\theta$ on the outage performance. We observe that there exists a unique $\theta$ which gives the best outage or ergodic capacity performance. For all three schemes, we see a similar trends on the impact of $\theta$, i.e., when $\theta$ increases from zero to the optimal value, the performance improves; when $\theta$ exceeds the optimal values, the performance deteriorates gradually. This phenomenon is rather intuitive, since performance of dual-hop systems is limited by the weakest hop quality. Moreover, we see that the optimal $\theta$ is in general different for different schemes and performance metrics, as shown in Fig. \ref{fig:4a}, the MMSE/MRT scheme requires a smaller $\theta$ compared with the MRC/MRT scheme, and the capacity optimal $\theta$ is larger than the outage optimal $\theta$ for the MMSE/MRT scheme.

\subsection{Effect of Key System Parameters on the Optimal $\theta$}
\begin{figure}[ht]
  \centering
  \subfigure[$N=2$, $\rho_1=10\;{\mbox{dB}}$, $\rho_I=0\;{\mbox{dB}}$]{\label{fig:5a}\includegraphics[width=0.21\textwidth]{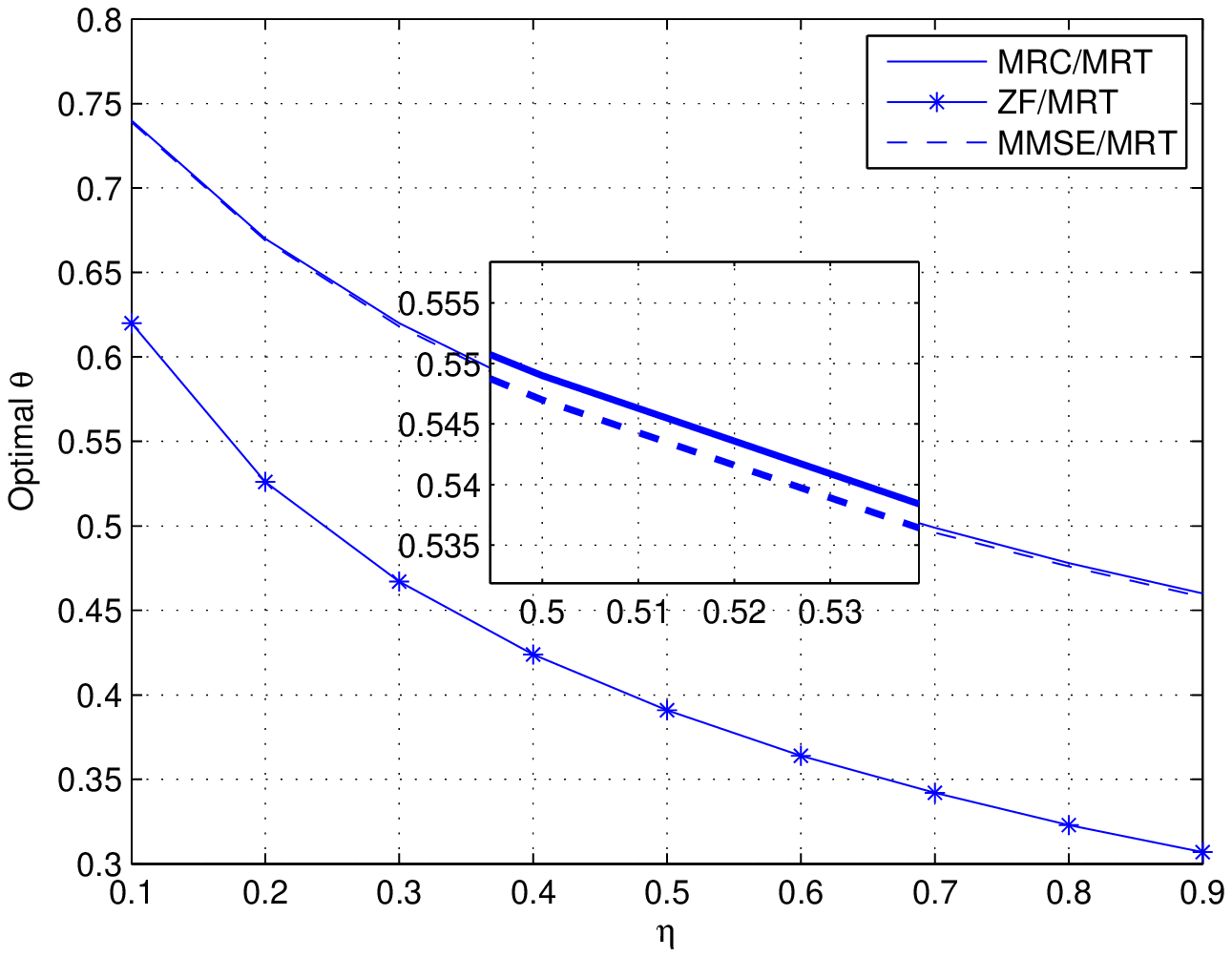}}
  \hspace{0.3in}
  \subfigure[$\eta=0.8$, $\rho_1=10\;{\mbox{dB}}$, $\rho_I=0\;{\mbox{dB}}$]{\label{fig:5b}\includegraphics[width=0.21\textwidth]{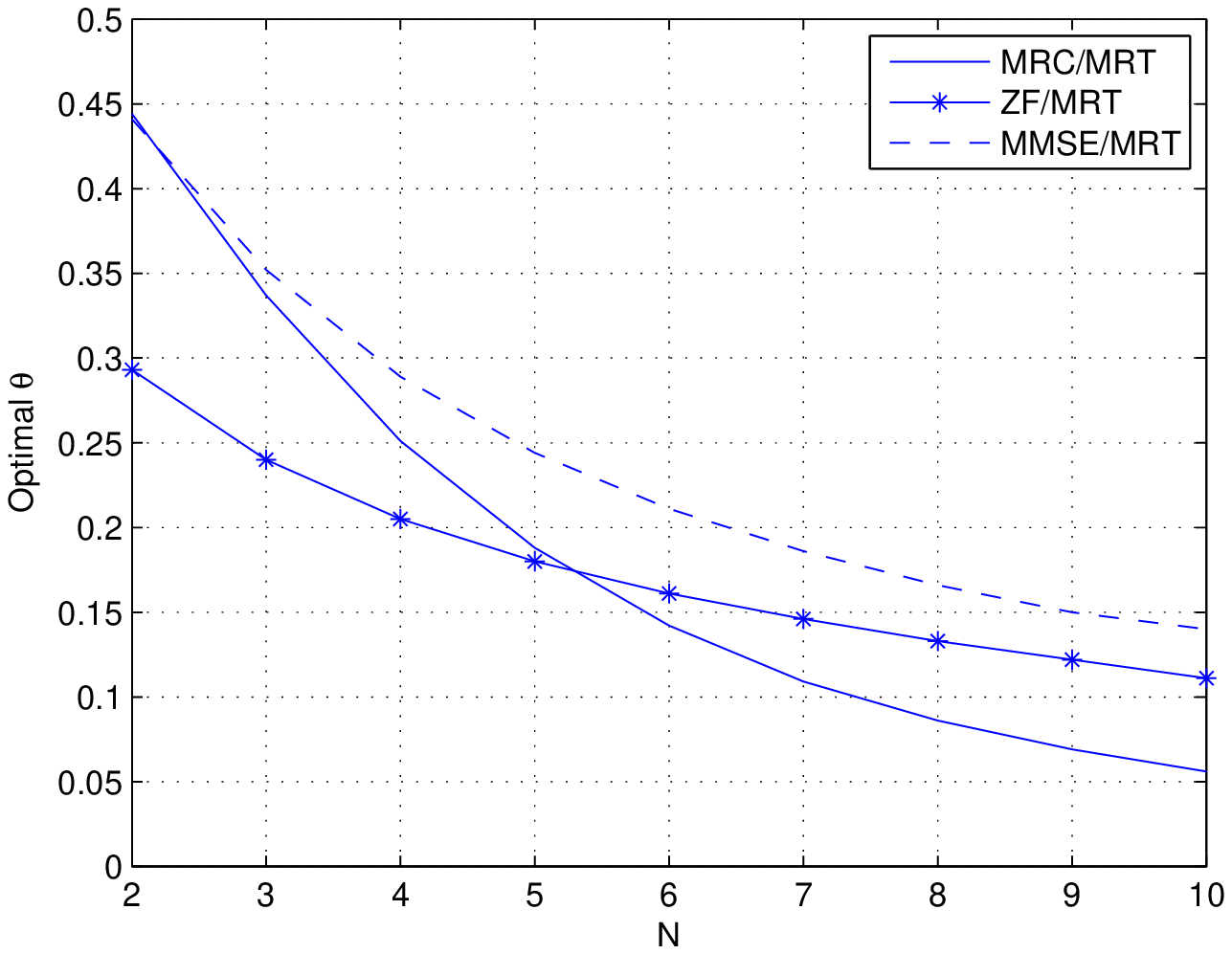}}
   \hspace{0.3in}
  \subfigure[$N=2$, $\eta=0.8$, $\rho_I=5\;{\mbox{dB}}$]{\label{fig:5c}\includegraphics[width=0.21\textwidth]{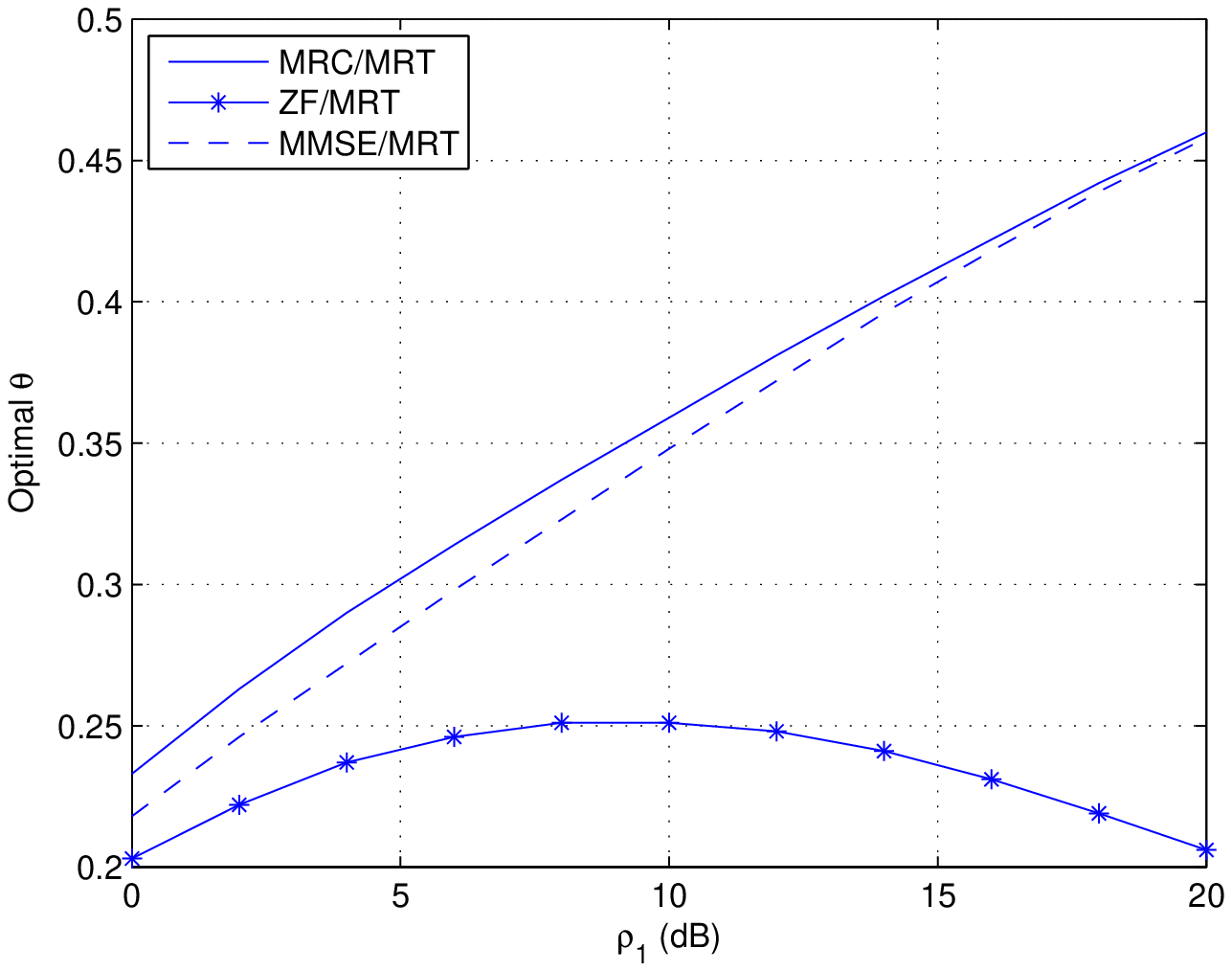}}
   \hspace{0.3in}
  \subfigure[$N=2$, $\eta=0.8$, $\rho_1=15\;{\mbox{dB}}$]{\label{fig:5d}\includegraphics[width=0.21\textwidth]{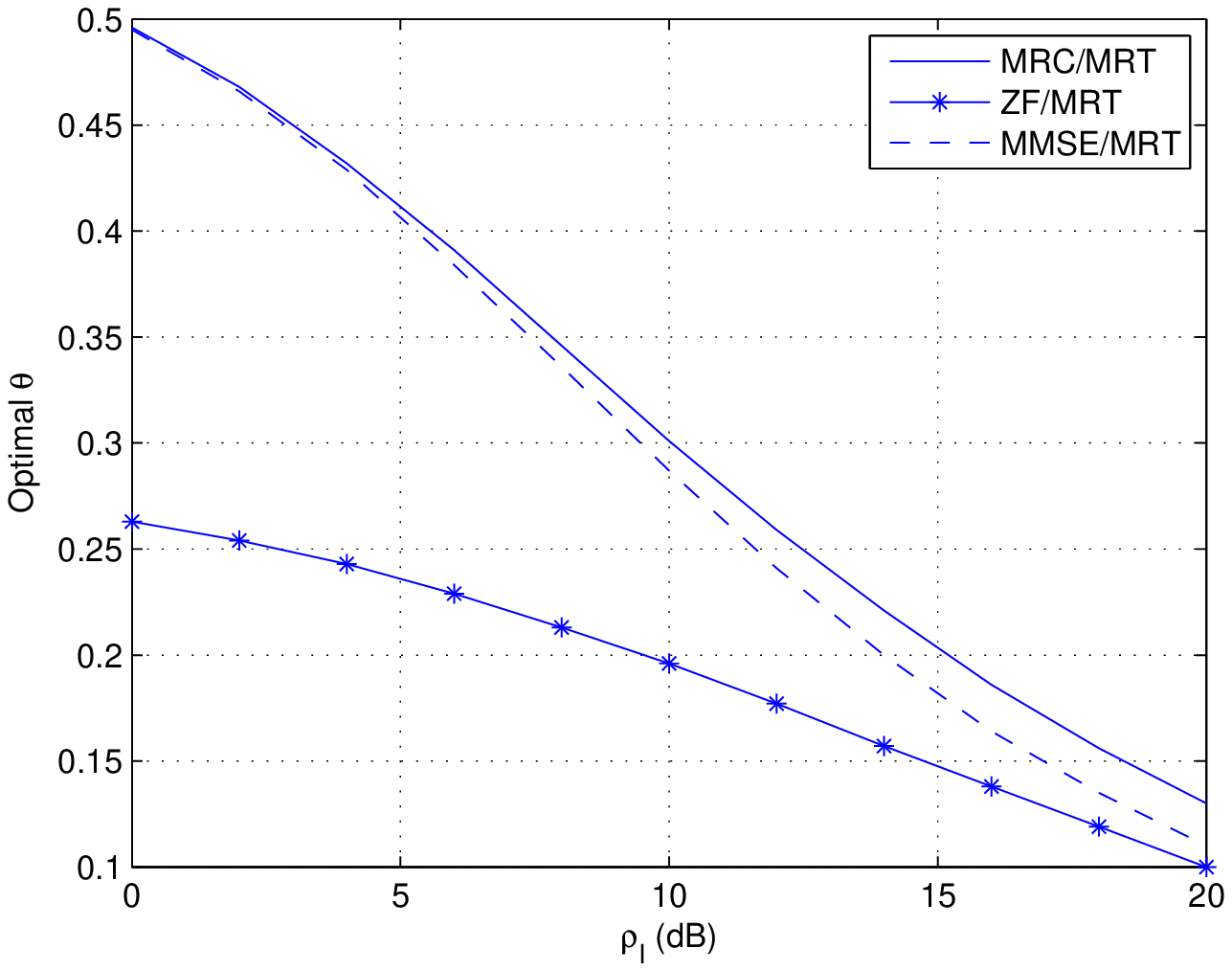}}
  \caption{The impact of (a) $\eta$, (b) $N$, (c) $\rho_1$, (d) $\rho_I$ on the outage optimal $\theta$.}
  \label{fig:fig5}
\end{figure}

Fig. \ref{fig:fig5} examines the effect of various key system parameters such as $\eta$, $N$, $\rho_1$ and $\rho_I$ on the choice of optimal $\theta$.  Specifically, Fig. \ref{fig:5a} illustrates the effect of $\eta$, and we can see that, the outage optimal $\theta$ is a decreasing function of $\eta$. A large $\eta$ implies higher energy conversion efficiency, which in turn suggests that less portion of the signal is needed for energy harvesting, hence, a smaller $\theta$ is required. A similar trend is observed in Fig. \ref{fig:5b} on the impact of $N$. As $N$ increases, the additional antennas improve the energy harvesting capability, i.e., more energy could be harvested, which implies that the optimal $\theta$ should decrease. The effect of $\rho_1$ is shown in Fig. \ref{fig:5c}, it is interesting to see that, for the MRC/MRT and MMSE/MRT schemes, the optimal $\theta$ is an increasing function of $\rho_1$, while for the ZF/MRT scheme, the optimal $\theta$ increases first along $\rho_1$ and decreases when $\rho_1$ exceeds certain value. Finally, Fig. \ref{fig:5d} investigates the effect of $\rho_I$. For all three schemes, the optimal $\theta$ is a decreasing function of $\rho_I$. This is intuitive since the CCI serves as the energy source, when the CCI power increases, a smaller $\theta$ is sufficient to fulfill the energy requirement at the relay. Similar trends could be observed for the capacity optimal $\theta$, which are not presented here due to space limitation.

\section{Conclusion}
 In this paper, we studied the performance of a dual-hop AF energy harvesting system with multiple antennas and CCI. Analytical expressions for the outage probability, ergodic capacity, as well as the diversity order were presented, which provide efficient means for the evaluation of the system's performance. In addition, the optimal power splitting ratio minimizing the outage probability was analytically characterized while the capacity optimal power splitting ratio was studied numerically. Moreover, the impact of various key system parameters, such as $\eta$, $N$, $\rho_1$ and $\rho_I$ on the optimal $\theta$ were examined, which provided useful design insights on the choice of a proper power splitting ratio under different system configurations.

 Our results demonstrate that both the MRC/MRT and MMSE/MRT schemes achieve a full diversity of $N$ while the ZF/MRT scheme only attains a diversity order of $N-1$. We showed that the CCI could be potentially exploited to significantly improve the system's performance. With the MMSE/MRT scheme, the CCI is always a desirable factor, and the stronger the CCI, the better the performance. Nevertheless, this is not the case for the MRC/MRT and ZF/MRT schemes, where the CCI could be detrimental. For instance, strong interference degrades the system performance of the MRC/MRT scheme, whereas the performance worsens in the presence of weak interference with the ZF/MRC scheme.

\appendices
\section{Proof of Corollary \ref{coro:1}}\label{appendix:corollary:1}
We first notice that the end-to-end SNR of the system can be tightly upper bounded by
\begin{align}\label{Aopa:5}
\gamma  < \frac{\frac{\eta \theta \left( {1 - \theta } \right)\rho _1^2}{d_1^{2\tau}d_2^{\tau}}\left\| {{{\bf{h}}_2}} \right\|_F^2\left\| {{{\bf{h}}_1}} \right\|_F^4}{\frac{\eta \theta {\rho _1}}{d_1^{\tau}d_2^{\tau}}\left\| {{{\bf{h}}_2}} \right\|_F^2\left\| {{{\bf{h}}_1}} \right\|_F^2 + \frac{\left( {1 - \theta } \right){\rho _1}}{d_1^{\tau}}\left\| {{{\bf{h}}_1}} \right\|_F^2}.
\end{align}
Hence, we get the following the outage probability lower bound:
\begin{align}\label{Aopa:6}
P_{{\sf{out}}}^{{\sf{low}}} &=  {\rm{Prob}}\left( {\left\| {{{\bf{h}}_2}} \right\|_F^2\left( {c\left\| {{{\bf{h}}_1}} \right\|_F^4 - d\left\| {{{\bf{h}}_1}} \right\|_F^2} \right) < a\left\| {{{\bf{h}}_1}} \right\|_F^2} \right),
\end{align}
which can be computed as
\begin{multline}\label{Aopa:7}
P_{{\sf{out}}}^{{\sf{low}}}
 = \int_0^{d/c} {{f_{\left\| {{{\bf{h}}_1}} \right\|_F^2}}\left( x \right)dx}  +\\ \int_{d/c}^\infty  {{f_{\left\| {{{\bf{h}}_1}} \right\|_F^2}}\left( x \right){F_{\left\| {{{\bf{h}}_2}} \right\|_F^2}}\left( {\frac{a}{{cx - d}}} \right)dx}.
\end{multline}
Noticing that $\left\| {{{\bf{h}}_1}} \right\|_F^2$ and $\left\| {{{\bf{h}}_2}} \right\|_F^2$ are i.i.d. gamma random variables, we have
\begin{align}\label{Aopa:8}
P_{{\sf{out}}}^{{\sf{low}}} = 1 - \int_{d/c}^\infty  {\frac{{\Gamma \left( {N,\frac{a}{{cx - d}}} \right)}}{{\Gamma \left( N \right)}}\frac{{{x^{N - 1}}}}{{\Gamma \left( N \right)}}{e^{ - x}}dx}.
\end{align}
Then, making a change of variable $cx - d = t$, (\ref{Aopa:8}) can be alternatively written as
\begin{align}\label{Aopa:9}
P_{{\sf{out}}}^{{\sf{low}}} = 1 - {\left( {\frac{1}{c}} \right)^N}{e^{ - d/c}}\int_0^\infty  {\frac{{\Gamma \left( {N,a/t} \right)}}{{\Gamma \left( N \right)}}{{\left( {t + d} \right)}^{N - 1}}{e^{ - t/c}}dt}.
\end{align}
Invoking the series expansion of incomplete gamma function \cite[Eq. (8.352.4)]{Tables} and applying the binomial expansion ${\left( {t + d} \right)^{N - 1}} = \sum\limits_{j = 0}^{N - 1} {{{N-1}\choose j}{t^j}{d^{N - j - 1}}} $, (\ref{Aopa:9}) can be further expressed as
\begin{multline}\label{Aopa:10}
P_{{\sf{out}}}^{{\sf{low}}} =  1 - {\left( {\frac{1}{c}} \right)^N}{e^{ - d/c}}\sum\limits_{i = 0}^{N - 1} {\frac{{{a^i}}}{{i!}}}\sum\limits_{j = 0}^{N - 1} {{{N-1}\choose j}} {d^{N - j - 1}}\\ \times \int_0^\infty  {{t^{j - i}}{e^{ - \left( {\frac{a}{t} + \frac{t}{c}} \right)}}dt} .
\end{multline}
To this end, with the help of \cite[Eq. (8.432.7)]{Tables}, the desired result can be obtained.


\section{Proof of Theorem \ref{theorem:2}}\label{appendix:theorem:2}
Starting from (\ref{SM:7}), we observe that, as ${\rho _1} \to \infty$, the end-to-end SNR can be tightly bounded by
\begin{align}\label{Aopa:11}
\gamma  < \gamma^{\sf up} = \min \left( {\frac{\left( {1 - \theta } \right){\rho _1}}{d_1^{\tau}}\left\| {{{\bf{h}}_1}} \right\|_F^2\;,\;\frac{\eta \theta {\rho _1}}{d_1^{\tau}d_2^{\tau}}\left\| {{{\bf{h}}_2}} \right\|_F^2\left\| {{{\bf{h}}_1}} \right\|_F^2} \right).
\end{align}
We now study the c.d.f. of $\gamma^{\sf up}$. Noticing that $\gamma^{\sf up} = \frac{\rho _1}{d_1^{\tau}}\left\| {{{\bf{h}}_1}} \right\|_F^2{Y}$, where $Y = \min \left( {\left( {1 - \theta } \right)\;,\;\frac{\eta \theta}{d_2^{\tau}} \left\| {{{\bf{h}}_2}} \right\|_F^2} \right)$, we first look at the c.d.f. of $Y$, which can be expressed as
\begin{multline}\label{Aopa:12}
{F_Y}\left( y \right) = \underbrace {{\rm{Prob}}\left( {\left\| {{{\bf{h}}_2}} \right\|_F^2 < \frac{yd_2^{\tau}}{{\eta \theta }}\;,\;\left\| {{{\bf{h}}_2}} \right\|_F^2 < \frac{(1 - \theta) d_2^{\tau}}{{\eta \theta }}} \right)}_{{{\rm{P}}_1}} + \\
\underbrace {{\rm{Prob}}\left( {1 - \theta  < y\;,\;\left\| {{{\bf{h}}_2}} \right\|_F^2 > \frac{(1 - \theta )d_2^{\tau}}{{\eta \theta }}} \right)}_{{{\rm{P}}_2}},
\end{multline}
with
\begin{align}\label{Aopa:13}
{{\rm{P}}_1} = \left\{ {\begin{array}{*{20}{c}}
{{\rm{Prob}}\left( {\left\| {{{\bf{h}}_2}} \right\|_F^2 < \frac{{(1 - \theta)d_2^{\tau} }}{{\eta \theta }}} \right)\;,\;y > 1 - \theta }\\
{{\rm{Prob}}\left( {\left\| {{{\bf{h}}_2}} \right\|_F^2 < \frac{yd_2^{\tau}}{{\eta \theta }}} \right)\;,\;\;\;\;y < 1 - \theta }
\end{array}} \right., \notag\\
\mbox{\quad}{{\rm{P}}_2} = \left\{ {\begin{array}{*{20}{c}}
{{\rm{Prob}}\left( {\left\| {{{\bf{h}}_2}} \right\|_F^2 > \frac{(1 - \theta )d_2^{\tau}}{{\eta \theta }}} \right)\;,\;\;y > 1 - \theta }\\
{0\;,\;\;\;\;\;\;\;\;\;\;\;\;\;\;\;\;\;\;\;\;\;\;\;\;\;\;\;\;\;\;\;\;\;y < 1 - \theta }
\end{array}} \right..
\end{align}

Therefore, the c.d.f. of $Y$ can be finally expressed as
\begin{align}\label{Aopa:15}
{F_Y}\left( y \right) &= \left\{ {\begin{array}{*{20}{c}}
{1\;,\;\;\;\;\;\;\;\;\;\;\;\;\;\;\;\;\;\;\;\;\;\;\;\;y > 1 - \theta }\\
{{\rm{Prob}}\left( {\left\| {{{\bf{h}}_2}} \right\|_F^2 < \frac{yd_2^{\tau}}{{\eta \theta }}} \right)\;,\;y < 1 - \theta \;\;}
\end{array}} \right. \notag\\
&= \;\left\{ {\begin{array}{*{20}{c}}
{1\;,\;\;\;\;\;\;\;\;\;\;\;\;\;\;\;\;\;\;\;\;y > 1 - \theta }\\
{1 - \frac{{\Gamma \left( {N,{yd_2^{\tau} \mathord{\left/
 {\vphantom {y {\eta \theta }}} \right.
 \kern-\nulldelimiterspace} {\eta \theta }}} \right)}}{{\Gamma \left( N \right)}}\;,\;y < 1 - \theta \;\;}
\end{array}} \right.
\end{align}

Having obtained the c.d.f. of $Y$, we are ready to compute the c.d.f. of $\gamma^{\sf up}$ as follows:
\begin{align}\label{Aopa:16}
{F_{\gamma^{\sf up}}}\left( z \right) = \int_0^\infty  {{F_Y}\left( {\frac{zd_1^{\tau}}{{{\rho _1}x}}} \right){f_{\left\| {{{\bf{h}}_1}} \right\|_F^2}}\left( x \right)dx},
\end{align}
which can be expressed as
\begin{align}\label{Aopa:17}
{F_{\gamma^{\sf up}}}\left( z \right)
 &= 1 - \int_{\frac{zd_1^{\tau}}{{\left( {1 - \theta } \right){\rho _1}}}}^\infty  {\frac{{\Gamma \left( {N,\frac{zd_1^{\tau}d_2^{\tau}}{{\eta \theta {\rho _1}x}}} \right)}}{{\Gamma \left( N \right)}}\frac{{{x^{N - 1}}{e^{ - x}}}}{{\Gamma \left( N \right)}}dx}.
\end{align}
%
Now, applying the asymptotic expansion of incomplete gamma function \cite[Eq. (8.354.2)]{Tables} to (\ref{Aopa:17}) yields
\begin{align}\label{Aopa:19}
{F_{\gamma^{\sf up}}}\left( z \right)  \approx 1 - \int_{\frac{zd_1^{\tau}}{{\left( {1 - \theta } \right){\rho _1}}}}^\infty  {\left( {1 - \frac{1}{{N!}}{{\left( {\frac{zd_1^{\tau}d_2^{\tau}}{{\eta \theta {\rho _1}x}}} \right)}^N}} \right)\frac{{{x^{N - 1}}{e^{ - x}}}}{{\Gamma \left( N \right)}}dx}.
\end{align}
Please note, due to the omission of the higher order items of the asymptotic expansion of incomplete gamma function, the expression given in (\ref{Aopa:19}) is no longer a bound, but a very tight asymptotic approximation, and matches {well with the exact value in the high SNR region, i.e., $P_{\sf out}^\infty \mathop  \to \limits_{{\rho _1} \to \infty } {\rm{Prob}}\left( {\gamma < \gamma_{\sf th}} \right)$.}

To this end, with the help of \cite[Eq. (8.350.2)]{Tables} and \cite[Eq. (8.211.1)]{Tables}, we obtain the following closed-form expression for $P_{\sf out}^\infty$
\begin{align}\label{Aopa:20}
P_{\sf out}^\infty = 1 - \frac{{\Gamma \left( {N,\frac{\gamma_{\sf th}d_1^{\tau}}{{\left( {1 - \theta } \right){\rho _1}}}} \right)}}{{\Gamma \left( N \right)}} - \frac{{{\rm{Ei}}\left( { - \frac{\gamma_{\sf th}d_1^{\tau}}{{\left( {1 - \theta } \right){\rho _1}}}} \right)}}{{N!\left( {N - 1} \right)!}}{\left( {\frac{\gamma_{\sf th}d_1^{\tau}d_2^{\tau}}{{\eta \theta {\rho _1}}}} \right)^N}.
\end{align}
Finally, applying \cite[Eq. (8.214.1)]{Tables} and \cite[Eq. (8.354.2)]{Tables} yields the desired result.
%
%

\section{Proof of Theorem \ref{theorem:3}}\label{appendix:theorem:3}
The ergodic capacity can be upper bounded by
\begin{align}\label{Aeca:1}
{C_{\sf up}} = {C_{{\gamma _1}}} + {C_{{\gamma _2}}} - \frac{1}{2}{\log _2}\left( {1 + {e^{{\mathop{\rm E}\nolimits} \left( {\ln {\gamma _1}} \right)}} + {e^{{\mathop{\rm E}\nolimits} \left( {\ln {\gamma _2}} \right)}}} \right).
\end{align}
Note that ${C_{{\gamma _1}}}$ is the ergodic capacity of the SIMO Rayleigh channel, which has been given in \cite{ZHU_CONF} as
\begin{align}\label{Aeca:4.2}
{C_{{\gamma _1}}} = \frac{1}{{2\ln 2}}\sum\limits_{k = 0}^{N - 1} {{{\left( {\frac{d_1^{\tau}}{{\left( {1 - \theta } \right){\rho _1}}}} \right)}^k}{e^{\frac{d_1^{\tau}}{{\left( {1 - \theta } \right){\rho _1}}}}}\Gamma \left( { - k,\frac{d_1^{\tau}}{{\left( {1 - \theta } \right){\rho _1}}}} \right)},
\end{align}
and ${C_{{\gamma _2}}}$ is the ergodic capacity of the SIMO keyhole channel, which has been given in \cite{H.Shin} as
\begin{align}\label{Aeca:10}
{C_{{\gamma _2}}} = \frac{1}{{2\ln 2}}\frac{1}{{\Gamma \left( {{N}} \right)}}\sum\limits_{m = 0}^{N - 1} {\frac{{\left( {\frac{d_1^{\tau}d_2^{\tau}}{{\eta \theta {\rho _1}}}} \right)}^m}{{m!}}G_{1,3}^{3,1}\left( \frac{d_1^{\tau}d_2^{\tau}}{{\eta \theta {\rho _1}}}\middle|^
{ - m}_
{ - m,N - m,0}\right)}.
\end{align}

Next, we observe that ${\mathop{\rm E}\nolimits} \left( {\ln {\gamma _1}} \right) = \ln \left( {\frac{\left( {1 - \theta } \right){\rho _1}}{d_1^{\tau}}} \right) + {\mathop{\rm E}\nolimits} \left( {\ln \left\| {{{\bf{h}}_1}} \right\|_F^2} \right)$ and ${\mathop{\rm E}\nolimits} \left( {\ln {\gamma _2}} \right) = \ln \left( {\frac{\eta \theta {\rho _1}}{d_1^{\tau}d_2^{\tau}}} \right) + {\mathop{\rm E}\nolimits} \left( {\ln \left\| {{{\bf{h}}_1}} \right\|_F^2} \right) + {\mathop{\rm E}\nolimits} \left( {\ln \left\| {{{\bf{h}}_2}} \right\|_F^2} \right)$. It is easy to show that
\begin{align}\label{Aeca:13}
{\mathop{\rm E}\nolimits} \left( {\ln \left\| {{{\bf{h}}_1}} \right\|_F^2} \right) = {\mathop{\rm E}\nolimits} \left( {\ln \left\| {{{\bf{h}}_2}} \right\|_F^2} \right) = \psi \left( N \right).
\end{align}

To this end, pulling everything together yeilds the desired result.


\section{Proof of Theorem \ref{theorem:4}}\label{appendix:theorem:4}
\begin{figure}[ht]
  \centering
  \subfigure[Probability versus $\rho_1$ in dB.]{\label{fig:8a}\includegraphics[width=0.45\textwidth]{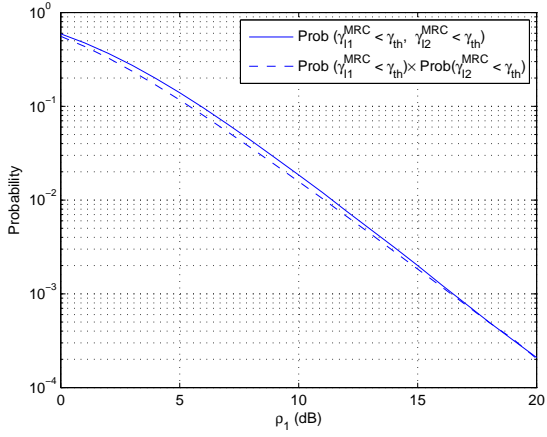}}
  \hspace{0.2in}
  \subfigure[Probability versus $\rho_1$ in dB.]{\label{fig:8b}\includegraphics[width=0.45\textwidth]{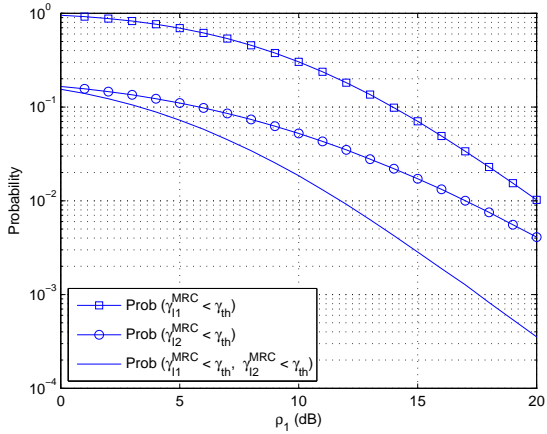}}
  \caption{Justification of the approximations employed in the proof of Theorem 4 and 5.}
  \label{fig:fig8}
\end{figure}

From (\ref{EOP:2}), the outage lower bound can be evaluated as
\begin{align}\label{AEOP:1}
P_{I\sf out}^{\sf LMRC} &= {\rm{Prob}}\left( {{\gamma _{I1}^{\sf MRC}} < {\gamma _{{\sf{th}}}}} \right) + {\rm{Prob}}\left( {{\gamma _{I2}^{\sf MRC}} < {\gamma _{{\sf{th}}}}} \right) \notag\\&-
{\rm{Prob}}\left( {{\gamma _{I1}^{\sf MRC}} < {\gamma _{{\sf{th}}}}\;\text{and} \;{\gamma _{I2}^{\sf MRC}} < {\gamma _{{\sf{th}}}}} \right)\notag\\
&\approx {\rm{Prob}}\left( {{\gamma _{I1}^{\sf MRC}} < {\gamma _{{\sf{th}}}}} \right) + {\rm{Prob}}\left( {{\gamma _{I2}^{\sf MRC}} < {\gamma _{{\sf{th}}}}} \right) \notag\\&- {\rm{Prob}}\left( {{\gamma _{I1}^{\sf MRC}} < {\gamma _{{\sf{th}}}}} \right){\rm{Prob}}\left( {{\gamma _{I2}^{\sf MRC}} < {\gamma _{{\sf{th}}}}} \right).
\end{align}
In general, $\gamma _{I1}^{\sf MRC}$ and $\gamma _{I2}^{\sf MRC}$ are not independent. However, through Monte Carlo simulations, {we observe that as long as $\rho_I$ is close to $\rho_1$, the term ${\rm{Prob}}\left( {{\gamma _{I1}^{\sf MRC}} < {\gamma _{{\sf{th}}}}\;\text{and} \;{\gamma _{I2}^{\sf MRC}} < {\gamma _{{\sf{th}}}}} \right)$ can be safely approximated by ${\rm{Prob}}\left( {{\gamma _{I1}^{\sf MRC}} < {\gamma _{{\sf{th}}}}} \right){\rm{Prob}}\left( {{\gamma _{I2}^{\sf MRC}} < {\gamma _{{\sf{th}}}}} \right)$ in the whole SNR region as shown in Fig \ref{fig:8a}}.
As a matter of fact, the same approximation has been adopted in \cite{Ikki}. Therefore, the remaining task is to compute ${\rm{Prob}}\left( {{\gamma _{I1}^{\sf MRC}} < {\gamma _{{\sf{th}}}}} \right)$ and ${\rm{Prob}}\left( {{\gamma _{I2}^{\sf MRC}} < {\gamma _{{\sf{th}}}}} \right)$.

The c.d.f. of ${\gamma_{I1}^{\sf MRC}}$ can be expressed as
\begin{align}\label{AEOP:2}
F_{\gamma_{I1}^{\sf MRC}}\left({\gamma_{\sf th}}\right)  = {\rm{Prob}}\left( {\left\| {{{\bf{h}}_1}} \right\|_F^2 < \frac{\left( {U + 1} \right)d_1^{\tau}{\gamma _{{\sf{th}}}}}{{\left( {1 - \theta } \right){\rho _1}}}} \right),
\end{align}
where $U = \frac{\left( {1 - \theta } \right){\rho _I}}{d_I^{\tau}}\frac{{{{\left| {{\bf{h}}_1^\dag {{\bf{h}}_I}} \right|}^2}}}{{\left\| {{{\bf{h}}_1}} \right\|_F^2}}$, which is an exponential random variables with parameter $\frac{\left( {1 - \theta } \right){\rho _I}}{d_I^{\tau}}$ \cite{H.Ding}.
Hence, we have
\begin{multline}
F_{\gamma_{I1}^{\sf MRC}}\left({\gamma_{\sf th}}\right)
 = 1 - \frac{{{d_I^{\tau}e^{ - \frac{{{d_1^{\tau}\gamma _{{\sf{th}}}}}}{{\left( {1 - \theta } \right){\rho _1}}}}}}}{{\left( {1 - \theta } \right){\rho _I}}}{\sum\limits_{m = 0}^{N - 1} {\frac{1}{{m!}}\left( {\frac{d_1^{\tau}\gamma _{{\sf{th}}}}{{\left( {1 - \theta } \right){\rho _1}}}} \right)} ^m}\\ \times\int_0^\infty  {{{\left( {x + 1} \right)}^m}{e^{ - \left( {\frac{d_1^{\tau}\gamma _{{\sf{th}}}}{{\left( {1 - \theta } \right){\rho _1}}} + \frac{d_I^{\tau}}{{\left( {1 - \theta } \right){\rho _I}}}} \right)x}}dx}.\label{AEOP:3b}
\end{multline}
%
Then, applying the binomial expansion and invoking \cite[Eq. (8.312.2)]{Tables}, we arrive at
\begin{multline}\label{AEOP:4}
F_{\gamma_{I1}^{\sf MRC}}\left({\gamma_{\sf th}}\right) = 1 - \frac{{{d_I^{\tau}e^{ - \frac{{{d_1^{\tau}\gamma _{{\sf{th}}}}}}{{\left( {1 - \theta } \right){\rho _1}}}}}}}{{\left( {1 - \theta } \right){\rho _I}}}{\sum\limits_{m = 0}^{N - 1} {\left( {\frac{{{d_1^{\tau}\gamma _{{\sf{th}}}}}}{{\left( {1 - \theta } \right){\rho _1}}}} \right)} ^m}\\
\times\sum\limits_{n = 0}^m {\frac{1}{{(m - n)!}}{{\left( {\frac{{\left( {1 - \theta } \right){\rho _1}{\rho _I}}}{{{d_I^{\tau}\rho _1} + {d_1^{\tau}\rho _I}{\gamma _{{\sf{th}}}}}}} \right)}^{n + 1}}}.
\end{multline}

Similarly, the c.d.f. of ${\gamma_{I2}^{\sf MRC}}$ can be expressed as
\begin{align}\label{AEOP:5}
F_{\gamma_{I2}^{\sf MRC}}\left({\gamma_{\sf th}}\right) &= {\rm{Prob}}\left( {{\gamma _{I2}^{\sf MRC}} < {\gamma _{{\sf{th}}}}} \right) ={\rm{Prob}}\left( {\left\| {{{\bf{h}}_2}} \right\|_F^2 < \frac{{{d_2^{\tau}\gamma _{{\sf{th}}}}}}{{\eta \theta Z}}} \right) \notag\\
&=\int_0^\infty  {{F_{\left\| {{{\bf{h}}_2}} \right\|_F^2}}\left( {\frac{{{d_2^{\tau}\gamma _{{\sf{th}}}}}}{{\eta \theta x}}} \right){f_Z}\left( x \right)dx},
\end{align}
where $Z = \frac{\rho _1}{d_1^{\tau}}\left\| {{{\bf{h}}_1}} \right\|_F^2 + \frac{\rho _I}{d_I^{\tau}}\left\| {{{\bf{h}}_I}} \right\|_F^2$, which is a sum of two independent gamma random variables. According to \cite{A.Abu-Dayya}, if ${\rho _1} \ne {\rho _I}$ the probability density function (p.d.f.) of $Z$ can be given by
\begin{multline}\label{AEOP:6}
{f_Z}(x) = \frac{d_1^{N\tau}d_I^{N\tau}}{{\rho _1^N\rho _I^N}}\sum\limits_{s = 1}^N {\frac{{\prod\nolimits_{j = 1}^{s - 1} {\left( {1 - N - j} \right)} }}{{\left( {N - s} \right)!\left( {s - 1} \right)!}}}\times\\{{\left( {\frac{d_I^{\tau}}{{{\rho _I}}} - \frac{d_1^{\tau}}{{{\rho _1}}}} \right)}^{1 - N - s}}{x^{N - s}}{e^{ - \frac{d_1^{\tau}x}{{{\rho _1}}}}}
 + \frac{d_1^{N\tau}d_I^{N\tau}}{{\rho _1^N\rho _I^N}}\times\\\sum\limits_{s = 1}^N {\frac{{\prod\nolimits_{j = 1}^{s - 1} {\left( {1 - N - j} \right)} }}{{\left( {N - s} \right)!\left( {s - 1} \right)!}}{{\left( {\frac{d_1^{\tau}}{{{\rho _1}}} - \frac{d_I^{\tau}}{{{\rho _I}}}} \right)}^{1 - N - s}}{x^{N - s}}{e^{ - \frac{d_I^{\tau}x}{{{\rho _I}}}}}}.
\end{multline}
After some algebraic manipulations and with the help of \cite[Eq. ( 8.432.7)]{Tables}, (\ref{AEOP:5}) can be computed as (\ref{AEOP:8}) shown on the top of the next page.
\begin{figure*}
\begin{multline}\label{AEOP:8}
F_{\gamma_{I2}^{\sf MRC}}\left({\gamma_{\sf th}}\right) = 1 - \frac{2d_1^{N\tau}d_I^{N\tau}}{{\rho _1^N\rho _I^N}}\sum\limits_{s = 1}^N {\frac{{\prod\nolimits_{j = 1}^{s - 1} {\left( {1 - N - j} \right)} }}{{\left( {N - s} \right)!\left( {s - 1} \right)!}}{{\left( {\frac{d_I^{\tau}}{{{\rho _I}}} - \frac{d_1^{\tau}}{{{\rho _1}}}} \right)}^{1 - N - s}}}
\sum\limits_{m = 0}^{N - 1} \frac{1}{{m!}}{{\left( {\frac{{{d_2^{\tau}\gamma _{\sf th}}}}{{\eta \theta }}} \right)}^{N + 1 - s}}\times \\{{\left( {\frac{{{d_1^{\tau}d_2^{\tau}\gamma _{\sf th}}}}{{\eta \theta {\rho _1}}}} \right)}^{\frac{{m + s - N - 1}}{2}}} {K_{m + s - N - 1}}\left( {2\sqrt {\frac{{{d_1^{\tau}d_2^{\tau}\gamma _{\sf th}}}}{{\eta \theta {\rho _1}}}} } \right)   +\frac{2d_1^{N\tau}d_I^{N\tau}}{{\rho _1^N\rho _I^N}}\sum\limits_{s = 1}^N {\frac{{\prod\nolimits_{j = 1}^{s - 1} {\left( {1 - N - j} \right)} }}{{\left( {N - s} \right)!\left( {s - 1} \right)!}}}  \times \\{{\left( {\frac{d_1^{\tau}}{{{\rho _1}}} - \frac{d_I^{\tau}}{{{\rho _I}}}} \right)}^{1 - N - s}}
\sum\limits_{m = 0}^{N - 1} {\frac{1}{{m!}}{{\left( {\frac{{d_2^{\tau}\gamma _{\sf th}}}{{\eta \theta }}} \right)}^{N + 1 - s}}{{\left( {\frac{{{d_2^{\tau}d_I^{\tau}\gamma _{\sf th}}}}{{\eta \theta {\rho _I}}}} \right)}^{\frac{{m + s - N - 1}}{2}}}{K_{m + s - N - 1}}\left( {2\sqrt {\frac{{{d_2^{\tau}d_I^{\tau}\gamma _{\sf th}}}}{{\eta \theta {\rho _I}}}} } \right)}.
\end{multline}
\hrule
\end{figure*}

To this end, substituting (\ref{AEOP:4}) and (\ref{AEOP:8}) into (\ref{AEOP:1}) yields the desired result.

\section{Proof of Theorem \ref{theorem:5}}\label{appendix:theorem:5}
When ${\rho _1} \to \infty $ the outage probability of the system can be approximated as
\begin{align}\label{AEOP:9}
{P_{I\sf out}^{\sf MRC}} \approx {\rm{Prob}}\left( {{\gamma _{I1}^{\sf MRC}} < {\gamma _{{\sf{th}}}}} \right) + {\rm{Prob}}\left( {{\gamma _{I2}^{\sf MRC}} < {\gamma _{{\sf{th}}}}} \right).
\end{align}
This approximation comes from the fact that, as $\rho_1$ increases, ${\rm{Prob}}\left( {{\gamma _{I1}^{\sf MRC}} < {\gamma _{{\sf{th}}}}\;\text{and} \;{\gamma _{I2}^{\sf MRC}} < {\gamma _{{\sf{th}}}}} \right)$  is negligible compared with ${\rm{Prob}}\left( {{\gamma _{I1}^{\sf MRC}} < {\gamma _{{\sf{th}}}}} \right)$ or ${\rm{Prob}}\left( {{\gamma _{I2}^{\sf MRC}} < {\gamma _{{\sf{th}}}}} \right)$ ({it can be justified through Fig \ref{fig:8b}}).

Therefore, the high SNR approximation for the outage probability can be given by
\begin{align}\label{AEOP:9.2}
P_{I\sf out}^{\sf MRC} \approx F_{\gamma_{I1}^{\sf MRC}}^\infty\left({\gamma_{\sf th}}\right) + F_{\gamma_{I2}^{\sf MRC}}^\infty\left({\gamma_{\sf th}}\right),
\end{align}
where $F_{\gamma_{I1}^{\sf MRC}}^\infty$ and $F_{\gamma_{I2}^{\sf MRC}}^\infty$ denote the high SNR approximation of ${\rm{Prob}}\left( {{\gamma _{I1}^{\sf MRC}} < {\gamma _{{\sf{th}}}}} \right)$ and ${\rm{Prob}}\left( {{\gamma _{I2}^{\sf MRC}} < {\gamma _{{\sf{th}}}}} \right)$, respectively.

We start with the characterization of $F_{\gamma_{I1}^{\sf MRC}}^\infty$. Starting from (\ref{AEOP:2}), and with the help of the asymptotic expansion of incomplete gamma function, we have
\begin{align}\label{AEOP:11}
F_{\gamma^{\sf MRC}_{I1}}^\infty\left({\gamma_{\sf th}}\right) = {\left( {\frac{{{d_1^{\tau}\gamma _{{\sf{th}}}}}}{{\left( {1 - \theta } \right){\rho _1}}}} \right)^N}\sum\limits_{n = 0}^N {\frac{{{{\left( {\left( {1 - \theta } \right){\rho _I}} \right)}^n}}}{{d_I^{n\tau}\left( {N - n} \right)!}}}.
\end{align}

Now, we turn our attention to $F_{\gamma_{I2}^{\sf MRC}}^\infty\left({\gamma_{\sf th}}\right)$.
According to (\ref{AEOP:5}) and utilizing the the asymptotic expansion of incomplete gamma function, conditioned on $y_1 =  \frac{\rho _1}{d_1^{\tau}}\left\| {{{\bf{h}}_1}} \right\|_F^2$ and $y_I =  \frac{\rho _I}{d_I^{\tau}}\left\| {{{\bf{h}}_I}} \right\|_F^2$, $F_{\gamma_{I2}^{\sf MRC}}^\infty\left({\gamma_{\sf th}}\right)$ can be expressed as
\begin{align}\label{AEOP:12}
F_{\gamma_{I2}^{\sf MRC}}^\infty\left({\gamma_{\sf th}}\right) = \frac{1}{{\Gamma \left( {N + 1} \right)}}{\left( {\frac{{{d_2^{\tau}\gamma _{{\sf{th}}}}}}{{\eta \theta \left( {{y_1} + {y_I}} \right)}}} \right)^N}.
\end{align}
Averaging over ${y_I}$, we have
\begin{multline}\label{AEOP:13}
F_{\gamma_{I2}^{\sf MRC}}^\infty\left({\gamma_{\sf th}}\right) = \frac{1}{{\Gamma \left( {N + 1} \right)\Gamma \left( N \right)}}{\left( {\frac{{{d_I^{\tau}d_2^{\tau}\gamma _{{\sf{th}}}}}}{{\eta \theta {\rho _I}}}} \right)^N}\times\\ \int_0^\infty  {{{\left( {\frac{1}{{{y_1} + x}}} \right)}^N}{x^{N - 1}}{e^{ - \frac{xd_I^{\tau}}{{{\rho _I}}}}}dx}.
\end{multline}
Make a change of variable ${y_1} + x = t$, and apply the binomial expansion, (\ref{AEOP:13}) can be rewritten by
\begin{multline}\label{AEOP:14}
F_{\gamma_{I2}^{\sf MRC}}^\infty\left({\gamma_{\sf th}}\right)
 = \frac{{{e^{\frac{{{d_I^{\tau}y_1}}}{{{\rho _I}}}}{\left( {\frac{{{d_2^{\tau}\gamma _{{\sf{th}}}}}}{{\eta \theta }}} \right)^N}}}}{{\Gamma \left( {N + 1} \right)\Gamma \left( N \right)}}\sum\limits_{i = 0}^{N - 1} {{N-1 \choose i}}{{\left( { - {y_1}} \right)}^{N - i - 1}} \\ \times \left({\frac{\rho _I}{d_I^{\tau}}}\right)^{i - 2N + 1}\Gamma \left( {i - N + 1,\frac{{{d_I^{\tau}y_1}}}{{{\rho _I}}}} \right).
\end{multline}
Further averaging over ${y_1}$, we have
\begin{multline}\label{AEOP:15}
F_{\gamma_{I2}^{\sf MRC}}^\infty\left({\gamma_{\sf th}}\right) = \frac{1}{{\Gamma \left( {N + 1} \right)\Gamma {{\left( N \right)}^2}}}{\left( {\frac{{{d_1^{\tau}d_2^{\tau}\gamma _{{\sf{th}}}}}}{{\eta \theta {\rho _1}}}} \right)^N}\times\\
\sum\limits_{i = 0}^{N - 1} {{N-1\choose i}{{\left( { - 1} \right)}^{N - i - 1}}\left({\frac{\rho _I}{d_I^{\tau}}}\right)^{i - 2N + 1}}\times \\
\underbrace {\int_0^\infty  {{e^{ - \left( {\frac{d_1^{\tau}}{{{\rho _1}}} - \frac{d_I^{\tau}}{{{\rho _I}}}} \right)x}}{x^{2N - i - 2}}\Gamma \left( {i - N + 1,\frac{xd_I^{\tau}}{{{\rho _I}}}} \right)dx} }_{{\cal I}_1}.
\end{multline}

With the help of \cite[Eq. (6.455.1)]{Tables}, the integral ${{\cal I}_1}$ can be solved as
\begin{multline}\label{AEOP:16}
{{\cal I}_1} = \frac{{\left({\frac{\rho _I}{d_I^{\tau}}}\right)^{N - i - 1}\left({\frac{\rho _1}{d_1^{\tau}}}\right)^N\Gamma \left( N \right)}}{{2N - i - 1}}\times\\{}_2{F_1}\left( {1,N;2N - i;1 - \frac{{{d_I^{\tau}\rho _1}}}{{{d_1^{\tau}\rho _I}}}} \right).
\end{multline}

Then, utilizing \cite[Eq. (9.131.1)]{Tables}, we can obtain
%

\begin{multline}\label{AEOP:18}
F_{\gamma_{I2}^{\sf MRC}}^\infty(\gamma_{\sf th}) =
\frac{{\left( {\frac{{{d_1^{\tau}d_2^{\tau}\gamma _{{\rm{th}}}}}}{{\eta \theta {\rho _1}}}} \right)^N}}{{\Gamma (N + 1)\Gamma(N)}}
\sum\limits_{i = 0}^{N - 1} {{N-1\choose i}}\times \\
{{\left( { - 1} \right)}^{N - i - 1}}\frac{{{}_2{F_1}\left( {N,2N - i - 1;2N - i;1 - \frac{{{d_I^{\tau}\rho _I}}}{{{d_1^{\tau}\rho _1}}}} \right)}}{{2N - i - 1}}.
\end{multline}

To this end, substituting (\ref{AEOP:11}) and (\ref{AEOP:18}) into (\ref{AEOP:9.2}) yields the desired result.

\section{Proof of Theorem \ref{theorem:6}}\label{appendix:theorem:6}
Similar to the proof of Theorem \ref{theorem:3}, we note that the ergodic capacity upper bound can be computed as
\begin{multline}\label{AEEC:1.0}
{C_{I\sf up}^{\sf MRC}} = {C_{{\gamma _{I1}^{\sf MRC}}}} + {C_{{\gamma _{I2}^{\sf MRC}}}} - \\
\frac{1}{2}{\log _2}\left( {1 + {e^{{\mathop{\rm E}\nolimits} \left( {\ln {\gamma _{I1}^{\sf MRC}}} \right)}} + {e^{{\mathop{\rm E}\nolimits} \left( {\ln {\gamma _{I2}^{\sf MRC}}} \right)}}} \right),
\end{multline}
where $C_{\gamma _{Ii}^{\sf MRC}} = \frac{1}{2}{\mathop{\rm E}\nolimits} \left[ {{{\log }_2}\left( {1 + {\gamma _{Ii}^{\sf MRC}}} \right)} \right]$, for $k \in \{ 1,2\} $. Hence, the remaining task is to compute ${C_{{\gamma _{I1}^{\sf MRC}}}}$, ${C_{{\gamma _{I2}^{\sf MRC}}}}$, ${{\mathop{\rm E}\nolimits} \left( {\ln {\gamma _{I1}^{\sf MRC}}} \right)}$ and ${{\mathop{\rm E}\nolimits} \left( {\ln {\gamma _{I2}^{\sf MRC}}} \right)}$.

\subsection{Calculation of $C_{\gamma _{I1}^{\sf MRC}}$}
Utilizing the same method as in \cite{Himal22}
and invoking the c.d.f. of $\gamma _{I1}^{\sf MRC}$ given in (\ref{AEOP:4}), $C_{\gamma _{I1}^{\sf MRC}}$ can be computed as
\begin{multline}\label{AEEC:1}
{C_{{\gamma _{I1}^{\sf MRC}}}} = \frac{1}{{2\ln 2}}\sum\limits_{m = 0}^{N - 1} {{{\left( {\frac{d_1^{\tau}}{{\left( {1 - \theta } \right){\rho _1}}}} \right)}^m}\sum\limits_{n = 0}^m {\frac{{{{\left( {\left( {1 - \theta } \right){\rho _I}} \right)}^n}}}{{d_I^{n\tau}\left( {m - n} \right)!}}} }\times\\
\underbrace {\int_0^\infty  {\frac{{{e^{ - \frac{xd_1^{\tau}}{{\left( {1 - \theta } \right){\rho _1}}}}}{x^m}}}{{\left( {1 + x} \right)}}{{\left( {1 + \frac{{{d_1^{\tau}\rho _I}}}{{{d_I^{\tau}\rho _1}}}x} \right)}^{ - \left( {n + 1} \right)}}dx} }_{{\cal {I}}_2}.
\end{multline}
With the help of the identity ${\left( {1 + \beta x} \right)^{ - \alpha }} = \frac{1}{{\Gamma \left( \alpha  \right)}}{\mathop{\rm G}\nolimits} _{1,1}^{1,1}\left( {\beta x\left|_0^{1-\alpha}\right.} \right)$ and the integral formula \cite[Eq. (2.6.2)]{H-function}, ${\cal I}_2$ can be computed as
\begin{align}\label{AEEC:2}
{\cal I}_2
&= \frac{{\left( {\left( {1 - \theta } \right){\rho _1}} \right)}^{m + 1}}{{d_1^{(m+1)\tau}\Gamma \left( {n + 1} \right)}}{\mathop{\rm G}\nolimits} _{1,[1:1],0,[1:1]}^{1,1,1,1,1}\left(^{\frac{\left( {1 - \theta } \right){\rho _1}}{d_1^{\tau}}}_{\frac{\left( {1 - \theta } \right){\rho _I}}{d_I^{\tau}}} \middle| \substack{{m + 1}\\
{0; - n}\\
 - \\
{0;0}}
\right).
\end{align}

\subsection{Calculation of $C_{\gamma _{I2}^{\sf MRC}}$}
Similarly, with the help of the c.d.f. of $\gamma _{I2}^{\sf MRC}$ given in (\ref{AEOP:8}), $C_{\gamma _{I2}^{\sf MRC}}$ can be computed as (\ref{AEEC:5}) shown on the top of the next page,
\begin{figure*}
\begin{multline}\label{AEEC:5}
{C_{{\gamma _{I2}^{\sf MRC}}}} = \frac{d_1^{N\tau}d_I^{N\tau}}{{\rho _1^N\rho _I^N\ln 2}}\sum\limits_{s = 1}^N {\frac{{\prod\nolimits_{j = 1}^{s - 1} {\left( {1 - N - j} \right)} }}{{\left( {N - s} \right)!\left( {s - 1} \right)!}}{{\left( {\frac{d_I^{\tau}}{{{\rho _I}}} - \frac{d_1^{\tau}}{{{\rho _1}}}} \right)}^{1 - N - s}}\sum\limits_{m = 0}^{N - 1} {\frac{1}{{m!}}{{\left( {\frac{d_2^{\tau}}{{\eta \theta }}} \right)}^{N + 1 - s}}{{\cal I}_3}} } \\
 + \frac{d_1^{\tau}d_I^{\tau}}{{\rho _1^N\rho _I^N\ln 2}}\sum\limits_{s = 1}^N {\frac{{\prod\nolimits_{j = 1}^{s - 1} {\left( {1 - N - j} \right)} }}{{\left( {N - s} \right)!\left( {s - 1} \right)!}}{{\left( {\frac{d_1^{\tau}}{{{\rho _1}}} - \frac{d_I^{\tau}}{{{\rho _I}}}} \right)}^{1 - N - s}}\sum\limits_{m = 0}^{N - 1} {\frac{1}{{m!}}{{\left( {\frac{d_2^{\tau}}{{\eta \theta }}} \right)}^{N + 1 - s}}{{\cal I}_4}} },
\end{multline}
\hrule
\end{figure*}
where
\begin{multline}\label{AEEC:6}
{\cal I}_3 = \int_0^\infty  {\frac{{{x^{N + 1 - s}}}}{{1 + x}}{{\left( {\frac{d_1^{\tau}d_2^{\tau}x}{{\eta \theta {\rho _1}}}} \right)}^{\frac{{m + s - N - 1}}{2}}}}\times \\
{K_{m + s - N - 1}}\left( {2\sqrt {\frac{d_1^{\tau}d_2^{\tau}x}{{\eta \theta {\rho _1}}}} } \right)dx,
\end{multline}
and
\begin{multline}\label{AEEC:7}
{\cal I}_4 = \int_0^\infty  {\frac{{{x^{N + 1 - s}}}}{{1 + x}}{{\left( {\frac{d_2^{\tau}d_I^{\tau}x}{{\eta \theta {\rho _I}}}} \right)}^{\frac{{m + s - N - 1}}{2}}}}\times\\
{K_{m + s - N - 1}}\left( {2\sqrt {\frac{d_2^{\tau}d_I^{\tau}x}{{\eta \theta {\rho _I}}}} } \right)dx.
\end{multline}
Then, following the similar lines as in the Appendix \ref{appendix:theorem:3}, ${C_{{\gamma _{I2}^{\sf MRC}}}}$ can be expressed in closed-form.

\subsection{Calculation of  ${{\mathop{\rm E}\nolimits} \left( {\ln {\gamma _{I1}^{\sf MRC}}} \right)}$ }

Invoking the c.d.f. of $\gamma _{I1}^{\sf MRC}$, the general moment of $\gamma _{I1}^{\sf MRC}$ can be computed as
\begin{multline}\label{AEEC:11}
{\rm{E}}\left( {(\gamma _{I1}^{\sf MRC})^k} \right) = \sum\limits_{m = 0}^{N - 1} {{{\left( {\frac{1}{{\left( {1 - \theta } \right){\rho _1}}}} \right)}^m}\sum\limits_{n = 0}^m {\frac{{{{\left( {\left( {1 - \theta } \right){\rho _I}} \right)}^n}}}{{\left( {m - n} \right)!}}k }}\times \\
\underbrace {\int_0^\infty  {{e^{ - \frac{x}{{\left( {1 - \theta } \right){\rho _1}}}}}{x^{m + k - 1}}{{\left( {1 + \frac{{{\rho _I}}}{{{\rho _1}}}x} \right)}^{ - \left( {n + 1} \right)}}dx} }_{{{\cal I}_5}}.
\end{multline}
With the help of \cite[Eq. (9.211.4)]{Tables}, we obtain
\begin{multline}\label{AEEC:13}
{\rm{E}}\left( {(\gamma _{I1}^{\sf MRC})^k} \right) = \sum\limits_{m = 0}^{N - 1} {{{\left( {\frac{d_1^{\tau}}{{\left( {1 - \theta } \right){\rho _1}}}} \right)}^m}\sum\limits_{n = 0}^m {\frac{{{{\left( {\left( {1 - \theta } \right){\rho _I}} \right)}^n}}}{{d_I^{n\tau}\left( {m - n} \right)!}}} }  \times \\
k{\left( {\frac{{{d_I^{\tau}\rho _1}}}{{{d_1^{\tau}\rho _I}}}} \right)^{m + k}}\Gamma ({m + k})\times\\
\Psi \left( {m + k,m + k - n;\frac{d_I^{\tau}}{{({1 - \theta }){\rho _I}}}} \right).
\end{multline}
The expectation of $\ln {\gamma _{I1}^{\sf MRC}}$ can be derived using ${\mathop{\rm E}\nolimits} \left( {\ln x} \right) = {\left. {\frac{{d{\mathop{\rm E}\nolimits} \left( {{x^k}} \right)}}{{dk}}} \right|_{k = 0}}$.
To proceed, we find it convenient to use (\ref{AEEC:14}) (shown on the top of the next page) as an alternative expression of (\ref{AEEC:13})
\begin{figure*}
\begin{multline}\label{AEEC:14}
{\rm{E}}\left( {(\gamma _{I1}^{\sf MRC})^k} \right) = \underbrace {{e^{\frac{d_I^{\tau}}{{\left( {1 - \theta } \right){\rho _I}}}}}\Gamma \left( {k + 1} \right){{\left( {\frac{{{d_I^{\tau}\rho _1}}}{{{d_1^{\tau}\rho _I}}}} \right)}^k}\Gamma \left( {1 - k,\frac{d_I^{\tau}}{{\left( {1 - \theta } \right){\rho _I}}}} \right)}_{{s_1}\left( k \right)} + \underbrace {\sum\limits_{m = 1}^{N - 1} {{{\left( {\frac{d_1^{\tau}}{{\left( {1 - \theta } \right){\rho _1}}}} \right)}^m}\sum\limits_{n = 0}^m {\frac{{{{\left( {(1 - \theta ){\rho _I}} \right)}^n}}}{{d_I^{n\tau}\left( {m - n} \right)!}}k{T_1}(k)} } }_{{s_2}\left( k \right)}.
\end{multline}
\hrule
\end{figure*}

where ${T_1}\left( k \right) = {\Psi \left( {m + k,m + k - n;\frac{d_I^{\tau}}{{\left( {1 - \theta } \right){\rho _I}}}} \right){\left( {\frac{{{d_I^{\tau}\rho _1}}}{{{d_1^{\tau}\rho _I}}}} \right)}^{m + k}}$ $\times \Gamma \left( {m + k} \right)$.
Now, the expectation of $\ln {\gamma _{I1}^{\sf MRC}}$ can be computed as
\begin{align}\label{AEEC:15}
{\mathop{\rm E}\nolimits} \left( {\ln {\gamma _{I1}^{\sf MRC}}} \right) = {\left. {\frac{{d{s_1}\left( k \right)}}{{dk}}} \right|_{k = 0}} + {\left. {\frac{{d{s_2}\left( k \right)}}{{dk}}} \right|_{k = 0}}.
\end{align}
It is easy to show that
\begin{multline}\label{AEEC:16}
{\left. {\frac{{d{s_1}\left( k \right)}}{{dk}}} \right|_{k = 0}} = \ln \left( {\left( {1 - \theta } \right){\rho _1}} \right)-\ln{d_1^{\tau}} + \psi \left( 1 \right) -\\
 {e^{\frac{d_I^{\tau}}{{\left( {1 - \theta } \right){\rho _I}}}}}G_{2,3}^{3,0}\left( \frac{d_I^{\tau}}{{\left( {1 - \theta } \right){\rho _I}}}\middle|^
{1,1}_
{0,0,1}
\right),
\end{multline}
where we have utilized the identity $\Gamma \left( {1,\frac{d_I^{\tau}}{{\left( {1 - \theta } \right){\rho _I}}}} \right) = {e^{ - \frac{d_I^{\tau}}{{\left( {1 - \theta } \right){\rho _I}}}}}$, and the derivative property $\frac{{d\Gamma \left( {a,z} \right)}}{{da}} = \Gamma \left( {a,z} \right)\ln z + G_{2,3}^{3,0}\left( z\middle|^{1,1}_{0,0,a}\right)$.

As for ${\left. {\frac{{d{s_2}\left( k \right)}}{{dk}}} \right|_{k = 0}}$, it is easy to observe that the key task is to compute ${\left. {\frac{{dk{T_1}\left( k \right)}}{{dk}}} \right|_{k = 0}}$. Hence, we have
\begin{align}\label{AEEC:17}
{\left. {\frac{{dk{T_1}\left( k \right)}}{{dk}}} \right|_{k = 0}} = {\left. {{T_1}\left( k \right)} \right|_{k = 0}} + k{\left. {\frac{{d{T_1}\left( k \right)}}{{dk}}} \right|_{k = 0}}.
\end{align}
Noticing that when $m \ge 1$, ${\left. {\frac{{d{T_1}\left( k \right)}}{{dk}}} \right|_{k = 0}} < \infty $ is a constant, we have
${\left. {k\frac{{d{T_1}\left( k \right)}}{{dk}}} \right|_{k = 0}} = 0$, hence ${\left. {\frac{{dk{T_1}\left( k \right)}}{{dk}}} \right|_{k = 0}} = {\left. {{T_1}\left( k \right)} \right|_{k = 0}}$. Therefore, we get
\begin{multline}\label{AEEC:18}
{\left. {\frac{{d{s_2}\left( k \right)}}{{dk}}} \right|_{k = 0}} = \sum\limits_{m = 1}^{N - 1} {\sum\limits_{n = 0}^m {\frac{{{{\left( {\left( {1 - \theta } \right){\rho _I}} \right)}^{n - m}}}}{{d_I^{(n-m)\tau}\left( {m - n} \right)!}}} }\times \\ \Gamma \left( m \right)\Psi \left( {m,m - n;\frac{d_I^{\tau}}{{\left( {1 - \theta } \right){\rho _I}}}} \right).
\end{multline}
To this end, substituting (\ref{AEEC:16}) and (\ref{AEEC:18}) into (\ref{AEEC:15}),  the expectation of $\ln {\gamma _{I1}^{\sf MRC}}$  can be obtained.

\subsection{Calculation of  ${{\mathop{\rm E}\nolimits} \left( {\ln {\gamma _{I2}^{\sf MRC}}} \right)}$ }
The expectation of ${{\mathop{\rm E}\nolimits} \left( {\ln {\gamma _{I2}^{\sf MRC}}} \right)}$ can be computed as
\begin{align}
{\mathop{\rm E}\nolimits} \left( {\ln {\gamma _{I2}^{\sf MRC}}} \right) = \ln \eta \theta -\ln{d_2^{\tau}} + {\mathop{\rm E}\nolimits} \left( {\ln Z} \right) + {\mathop{\rm E}\nolimits} \left( {\ln \left\| {{{\bf{h}}_2}} \right\|_F^2} \right).\notag
\end{align}
Since ${\mathop{\rm E}\nolimits} \left( {\ln \left\| {{{\bf{h}}_2}} \right\|_F^2} \right) = \psi \left( N \right)$, the remaining task is to figure out $ {\mathop{\rm E}\nolimits} \left( {\ln Z} \right)$. With the help of the p.d.f. of $Z$ given in (\ref{AEOP:6}), we have (\ref{AEEC:21}) shown on the top of the next page.

\begin{figure*}
\begin{multline}\label{AEEC:21}
{\mathop{\rm E}\nolimits} \left( {\ln Z} \right) = \frac{d_1^{N\tau}d_I^{N\tau}}{{\rho _1^N\rho _I^N}}\sum\limits_{s = 1}^N {\frac{{\prod\nolimits_{j = 1}^{s - 1} {\left( {1 - N - j} \right)} }}{{\left( {N - s} \right)!\left( {s - 1} \right)!}}{{\left( {\frac{d_I^{\tau}}{{{\rho _I}}} - \frac{d_1^{\tau}}{{{\rho _1}}}} \right)}^{1 - N - s}}\int_0^\infty  {\ln x\;{x^{N - s}}{e^{ - \frac{d_1^{\tau}x}{{{\rho _1}}}}}dx} } \\
 + \frac{d_1^{N\tau}d_I^{N\tau}}{{\rho _1^N\rho _I^N}}\sum\limits_{s = 1}^N {\frac{{\prod\nolimits_{j = 1}^{s - 1} {\left( {1 - N - j} \right)} }}{{\left( {N - s} \right)!\left( {s - 1} \right)!}}{{\left( {\frac{d_1^{\tau}}{{{\rho _1}}} - \frac{d_I^{\tau}}{{{\rho _I}}}} \right)}^{1 - N - s}}\int_0^\infty  {\ln x\;{x^{N - s}}{e^{ - \frac{d_I^{\tau}x}{{{\rho _I}}}}}dx} }.
\end{multline}
\hrule
\end{figure*}

Then, invoking \cite[Eq. (4.352.1)]{Tables}, the integral in (\ref{AEEC:21}) can be solved.

\nocite{*}
\bibliographystyle{IEEE}
\begin{footnotesize}

\end{footnotesize}

\end{document}